\documentclass[useAMS,usenatbib]{mn2e} 
\usepackage{graphics}
\usepackage{times}
\usepackage{amssymb}

\def\clover{{\sevensize C}$_\ell${\sevensize OVER}}
\def\wmap{{\sevensize WMAP}}

\def\quad{{\sevensize QUaD}}
\def\bicep{{\sevensize BICEP}}
\def\boomerang{{\sevensize BOOMERANG}}
\def\capmap{{\sevensize CAPMAP}}

\def\quiet{{\sevensize QUIET}}
\def\piper{{\sevensize PIPER}}
\def\polarbear{{\sevensize POLARBEAR}}
\def\ebex{{\sevensize EBEX}}
\def\spider{{\sevensize SPIDER}}
\def\cbi{{\sevensize CBI}}
\def\dasi{{\sevensize DASI}}
\def\brain{{\sevensize BRAIN}}
\def\planck{{\sevensize PLANCK}}
\def\maxipol{{\sevensize MAXIPOL}}
\def\cmbpolbpol{{\sevensize CMBPOL/BPOL}}

\def\healpix{{\sevensize HEALPIX}}
\def\camb{{\sevensize CAMB}}
\def\fftw{{\sevensize FFTW}}
\def\camgrid{{\sevensize CAMGRID}}

\def\be{\begin{equation}}
\def\ee{\end{equation}}
\def\ba{\begin{eqnarray}}
\def\ea{\end{eqnarray}}
\def\nn{\nonumber}
\def\lgl{\langle}
\def\rgl{\rangle}
\def\vnhat{\hat{\bmath{n}}}
\def\vx{\bmath{x}}

\def\valpha{\bmath{\alpha}}
\def\vbeta{\bmath{\beta}}
\def\vgrad{\bmath{\nabla}}
\def\vgamma{\bmath{\gamma}}
\def\mC{\mathbfss{C}}
\def\lsim{\mathrel{\rlap{\lower4pt\hbox{\hskip1pt$\sim$}}
    \raise1pt\hbox{$<$}}}                
\def\gsim{\mathrel{\rlap{\lower4pt\hbox{\hskip1pt$\sim$}}
    \raise1pt\hbox{$>$}}}                
\pagerange{\pageref{firstpage}--\pageref{lastpage}}
\pubyear{2009}

\begin{document}

\label{firstpage}

\title[Impact of modulation on CMB $B$-mode experiments]{Impact of
  modulation on CMB $\bmath B$-mode polarization experiments}
  \author[M.L. Brown et al.]{Michael L. Brown$^{1}$\thanks{E-mail: mbrown@mrao.cam.ac.uk}, 
  Anthony Challinor$^{2,3}$, Chris
  E. North$^{4}$\thanks{Current address: School of Physics and
  Astronomy, Cardiff University, Queen's
  Buildings, The Parade, Cardiff CF24 3A}, Bradley
  R. Johnson$^{4}$\thanks{Current address: Department of Physics, University of
  California, Berkeley, CA, 94720, USA},
  \newauthor Daniel O'Dea$^{1}$ and David Sutton$^{4}$\\ 
  $^{1}$ Astrophysics Group, Cavendish Laboratory, University of
  Cambridge, Cambridge CB3 OHE \\
  $^{2}$ Institute of Astronomy, University of Cambridge, Madingley
  Road, Cambridge CB3 OHA \\
  $^{3}$ DAMTP, Centre for Mathematical Sciences, University of
  Cambridge, Wilberforce Road, Cambridge CB3 OWA \\
  $^{4}$ Oxford Astrophysics, University of Oxford, Denys Wilkinson
  Building, 1 Keble Road, OX1 3RH}
\date{\today} 

\maketitle

\begin{abstract}
  We investigate the impact of both slow and fast polarization
  modulation strategies on the science return of upcoming ground-based
  experiments aimed at measuring the $B$-mode polarization of the
  cosmic microwave background. Using detailed simulations of 
  the \clover\, experiment, we compare the
  ability of modulated and un-modulated observations to recover the
  signature of gravitational waves in the polarized CMB sky in the
  presence of a number of anticipated systematic effects.  The general
  expectations that fast modulation is helpful in
  mitigating low-frequency detector noise, and that the additional
  redundancy in the projection of the instrument's polarization
  sensitivity directions onto the sky when modulating reduces the
  impact of instrumental polarization, particularly for fast
  modulation, are borne out by our simulations. Neither low-frequency
  polarized atmospheric fluctuations nor systematic errors in the
  polarization sensitivity directions are mitigated by
  modulation. Additionally, we find no significant reduction in the
  effect of pointing errors by modulation. For a \clover-like
  experiment, pointing jitter should be negligible but any systematic
  mis-calibration of the polarization coordinate reference system
  results in significant $E$-$B$ mixing on all angular scales and will
  require careful control.  We also stress the importance of combining
  data from multiple detectors in order to remove the effects of
  common-mode systematics (such as un-polarized $1/f$ atmospheric
  noise) on the measured polarization signal. Finally we compare the
  performance of our simulated experiment with the predicted
  performance from a Fisher analysis.  We find good agreement between
  the (optimal) Fisher predictions and the simulated experiment except
  for the very largest scales where the power spectrum estimator we
  have used introduces additional variance to the $B$-mode signal
  recovered from our simulations. In terms of detecting the total
  $B$-mode signal, including lensing, the Fisher analysis and the
  simulations are in excellent agreement. For a detection of the
  primordial $B$-mode signal only, using an input tensor-to-scalar 
  ratio of $r=0.026$, the Fisher analysis predictions are $\sim
  20$~per cent.
  better than the simulated performance.
\end{abstract}

\begin{keywords}
  methods: statistical - methods: analytical -
  cosmology: theory - cosmic microwave background 
\end{keywords}


\section{Introduction}
\label{sec:intro}
There is currently a great deal of interest in the rapidly evolving
field of observations of the polarization of the cosmic microwave
background (CMB). This interest stems from the fact that such
observations have the potential to discriminate between inflation and
other early-universe models through their ability to constrain an
odd-parity $B$-mode polarization component induced by a stochastic
background of gravitational waves at the time of last
scattering~\citep{kamionkowsky97,seljak97}.

From an observational point of view, we are still a long way off from a
detection of this $B$-mode signature of inflation. However, much
progress has been made recently with the detection of the much
stronger $E$-mode polarization signal on large scales by the \wmap\,
experiment \citep{page07,nolta08}.
On smaller scales, a growing number of
balloon-borne and ground-based experiments have also measured $E$-mode
polarization including \dasi\, \citep{leitch05}, \cbi\,
\citep{sievers07}, \boomerang\, \citep{montroy06}, \maxipol\,
\citep{wu07}, \capmap\, \citep{bischoff08} and \quad\,
\citep{pryke09}.  Most recently, the high precision measurement of
small-scale polarization by the \quad\, experiment has, for the first
time, revealed a characteristic series of acoustic peaks in the
$E$-mode spectrum and put the strongest upper limits to date on the
small-scale $B$-mode polarization signal expected from gravitational
lensing by large-scale structure.

Building on the experience gained from these pioneering experiments, a
new generation of experiments is now under
construction with the ambitious goal of observing the primordial
$B$-mode signal. Observing this signal is one of the most challenging goals
of modern observational cosmology. There are a number of reasons why
these types of observations are so difficult. First and foremost, the
sought-after signal is expected to be extremely small -- in terms of the 
tensor-to-scalar ratio\footnote{%
Our normalisation conventions follow those adopted in the
\camb\, code~\citep{lewis00}, so that $r$ is the ratio of primordial
power spectra for gravitational waves and curvature perturbations.
Explicitly, for slow-roll inflation in a potential $V(\phi)$,
$2 r\approx M_{\rm Pl}^2 [V'(\phi)/V(\phi)]^2$ where
$M_{\rm Pl}= 2.436 \times 10^{18}\, \mathrm{GeV}/c^2$ is the reduced
Planck mass.}, the RMS polarization signal from primordial $B$-modes is
$0.4 \sqrt{r}\, \mu \mathrm{K}$, and the current 95 per cent limit $r< 0.22$ from
\wmap\, temperature and $E$-mode polarization plus distance
indicators~\citep{komatsu09} implies an RMS $<180\,\mathrm{nK}$. Secondly,
polarized emission from
our own galaxy and from extra-Galactic objects act as a foreground
contaminant in observing the CMB polarized sky. Although our
knowledge of such polarized foregrounds is currently limited,
particularly at the higher frequencies $\gsim 100\,\mathrm{GHz}$
of relevance to bolometer experiments, models suggest that such contamination
could be an order of magnitude larger than the sought-after signal on
the largest scales (e.g. \citealt{amblard07}).
Thirdly, gravitational lensing by large-scale structure
converts $E$-modes into $B$-modes on small to medium scales
(see~\citealt{lewis06} for a review) and acts
as a source of confusion in attempts to measure the primordial
$B$-mode signal \citep{knox02, kesden02}. Note however that the
lensing $B$-mode signal is a valuable source of cosmological
information in its own right and can be used to put unique constraints
on dark energy and massive neutrinos
(e.g. \citealt{kaplinghat03,smithchallinor06}).
Unfortunately these latter two effects (foreground contamination and
weak gravitational lensing) contrive in such a way as to render $B$-mode
polarization observations subject to contamination on all angular
scales (the primordial $B$-mode signal is dominated by foregrounds on
large scales whilst on smaller scales it is swamped by the lensing
signal). Last, but not least, exquisite control of systematic and
instrumental effects will be required, to much better than $100\, \mathrm{nK}$,
before any detection of $B$-modes can be claimed.
The sought-after signal is so small that
systematic and instrumental effects considered negligible for an
exquisitely precise measurement of $E$-modes say, could potentially
ruin a detection of $B$-modes, if left uncorrected. One possible
approach to mitigating some of these systematics in hardware is to
modulate the incoming polarization signal such that it is shifted to
higher frequency and thus away from low-frequency systematics which
would otherwise contaminate it. There are a number of
techniques for achieving this including the use of a rotating
half-wave plate (HWP; e.g.~\citealt{johnson07}),
phase-switching, or Faraday rotation modulators~\citep{keating03}.

In this paper, we investigate the ability of 
modulation techniques that are either slow or fast 
with respect to the temporal variation of the
signal
to mitigate a range of possible systematic
effects. Our analysis is based on simulations, and the
subsequent analysis, of data from a ground-based CMB $B$-mode
polarization experiment. Previous investigations of the impact of
systematic effects on $B$-mode observations include the analytic works
of \citet{hu03}, \citet{odea07} and~\citet{shimon08},
as well as the simulation-based
analysis of \cite{mactavish08} who based their study on signal-only
simulations of the \spider\, experiment. The simulation work presented
here is complementary to these previous analyses but we also take our
analysis further by including realistic noise in our simulations ---
we are thus able to quantify not only any bias found, but also any
degradation of performance due to the presence of systematic
effects. Our work makes use of a detailed simulation pipeline which we
have created in the context of the \clover\, experiment. Although the
precise details of our simulations are specific to \clover\,, our
general conclusions regarding the impact of modulation on a variety of
systematic effects are relevant to all upcoming ground-based $B$-mode
experiments and many of them are also relevant for both balloon-borne
and space-based missions.

The paper is organised as follows. In Section \ref{sec:cmb_expts}, we
review the relevant upcoming $B$-mode polarization
experiments. Section \ref{sec:sims} describes our simulation technique
and the systematic effects we have considered. In Section
\ref{sec:analysis}, we describe the map-making and power spectrum
estimation techniques that we use to analyse the simulated
data. Section \ref{sec:results} presents the results from our main
analysis of systematic effects. We discuss our results in Section
\ref{sec:discussion} where, for clarity, we group the possible
systematic effects considered into those that are, and those that are
not, mitigated by a modulation scheme. In this section, we also
demonstrate the importance of combining information from multiple
detectors during analysis and compare the simulated performance of
\clover\, to the predicted performance from a Fisher
analysis. Our conclusions are summarised in Section
\ref{sec:conclusions}. Finally, Appendix~\ref{app:pointing} develops
a simple model of the spurious $B$-mode power produced by
pointing jitter for experiments with a highly redundant scan strategy.

\section{CMB polarization experiments}
\label{sec:cmb_expts}
A host of experiments are currently under construction (one is, in
fact, already observing) with their primary goal being to constrain the
tensor-to-scalar ratio, and thus the energy scale of inflation,
through observations of the $B$-mode component of the CMB. Here, we
give a short summary of the planned experiments.
\vspace{-3mm}
\subsubsection*{(i) Ground-based experiments:}
\begin{itemize}
  \item{{\sevensize BICEP/BICEP-2/KECK} array:} The \bicep\,
  experiment \citep{yoon06} has recently completed its third and final season of
  operation from the South Pole. The experiment consisted of a total of 98 polarization-sensitive
  bolometers (PSBs) at 100 and 150~GHz. The optical design is
  very clean but with the downside of poor resolution -- 45 arcmin
  full-width at half-maximum (FWHM) at 150~GHz -- which limits the
  target multipole range to $\ell <300$. The stated target sensitivity
  is $r = 0.1$. In the first observing season, three 100~GHz pixels and
  three 150~GHz pixels were equipped with Faraday rotation modulators
  and two pixels operated at 220~GHz. {\sevensize
  BICEP-2} will consist of an upgrade to the \bicep\, telescope with a
  150~GHz 512-element array of antenna-coupled detectors. It will be
  deployed to the South Pole in November, 2009. The {\sevensize KECK}
  array will consist of three {\sevensize BICEP-2}-like telescopes
  (at 100, 150 and 220~GHz). It is hoped to be installed on the
  \dasi\, mount (previously occupied by \quad) in November 2010. The
  nominal goal of this array is $r=0.01$.

  \vspace{3mm}\item{{\clover}:} For an up-to-date overview
  see~\citet{north08}. \clover\, is a
  three-frequency (97, 150 and 225~GHz) instrument to be sited at Pampa La
  Bola in the Atacama desert in Chile. It will have 576 single-polarization
  transition-edge sensors (TES), split equally among the three frequencies.
  The beam size of $\sim$ 5.5 arcmin FWHM at 150~GHz will sample the 
   multipole range $25 < \ell < 2000$. The target sensitivity is
  $r \sim 0.03$ and the polarization signal will be modulated with a
  HWP. The 97~GHz instrument is expected to be deployed to Chile in
  late 2009 with the combined 150/225~GHz instrument to follow soon
  after in 2010. 

  \vspace{3mm}\item{{\quiet}:} See~\citet{samtleben08} for a recent overview.
  \quiet\, is unique among
  planned $B$-mode experiments in that it uses pseudo-correlation
  HEMT-based receivers rather than bolometers. It will observe from Chile
  using the CBI mount -- a planned second phase will involve upgrading
  to $\sim 1000$ element arrays and relocation of the 7-m Crawford Hill
  antenna from New Jersey to Chile. \quiet\, will observe
  at 40 and 90~GHz. The target sensitivity for the second phase
  is $r \sim 0.01$.

  \vspace{3mm}\item{{\polarbear}:} A three-frequency (90, 150 and
    220~GHz) single-dish instrument to be sited in the Inyo Mountains, CA
  for its first year of operation, after which it will be relocated to
  the Atacama desert in Chile. It will use 1280 TES bolometers at each 
  frequency with polarization modulation from a HWP. The planned beam
  size is 4 arcmin (FWHM) at 150~GHz. The target sensitivity is
  $r=0.015$ for the full instrument.

  \vspace{3mm}\item{{\brain}:} See~\cite{charlassier08} for a recent
  review. \brain\, is a unique
  bolometric interferometer project (c.f. \dasi, \cbi) to be sited on
  the Dome-C site in Antarctica. The final instrument will have $\sim 1000$
  bolometers observing at 90, 150 and 220~GHz. \brain\, will be primarily
  sensitive to multipoles $50 < \ell < 200$. The full experiment is planned
  to be operational in 2011 and the stated target sensitivity is $r = 0.01$.
\end{itemize}
\vspace{-5mm}
\subsubsection*{(ii) Balloon-borne experiments}
\begin{itemize}
  \item{{\ebex}:} See \cite{oxley04} for a summary. \ebex\, will
  observe at 150, 250 and 410~GHz and will fly a total of
  $1406$ TES with HWP modulation.
  The angular resolution is 8 arcmin and
  the target multipole range is $20 < \ell < 2000$. The stated target
  sensitivity is $r = 0.02$.  A test
  flight is planned for 2009 and a long-duration balloon (LDB)
  flight is expected soon after.

  \vspace{3mm}\item{{\spider}:} See~\cite{crill08} for a recent description.
  \spider\, will deploy $\sim 3000$ antenna-coupled TES observing
  at 96, 145, 225 and 275~GHz, with a beam size of $\sim 40$ arcmin at 145~GHz.
  The target multipole range is $10 < \ell < 300$.  A 2-6 day first
  flight is planned for 2010. The target sensitivity is
  $r=0.01$. Signal modulation will be provided by a (slow) stepped HWP
  and fast gondola rotation.

  \vspace{3mm}\item{{\piper}:} This balloon experiment will deploy a
  focal plane of 5120 TES bolometers in a backshort-under-grid (BUG)
  configuration. Each flight of \piper\, will observe at a different
  frequency, covering 200, 270, 350 and 600~GHz after the four planned
  flights. The beam size is $\sim15$ arcmin, corresponding to a target
  multipole range $\ell < 800$.  The first element of the optical system is
  a variable polarization modulator (VPM).  The entire optical chain,
  including the modulators, are cooled to 1.5 K so that \piper\, observes
  at the background limit for balloon altitudes.  Including removal of
  foregrounds, the experiment has the sensitivity to make a $2\sigma$
  detection of $r = 0.007$. The first flight is scheduled for 2013.

\end{itemize}
\vspace{-5mm}
\subsubsection*{(iii) Space missions}
\begin{itemize}
  \item{{\planck}:} See the publication of the~\citet{planck06} for a detailed
   review of the science programme. \planck\, will measure the temperature
   in nine frequency bands, seven of which will have (some) polarization
   sensitivity. The polarized channels (100, 143, 217 and 353~GHz)
   of the high-frequency instrument
   (HFI) use similar PSBs to those deployed on
   \boomerang\, and have beam sizes 5--9.5 arcmin. For low $r$, sensitivity to
   primordial gravitational waves will
   mostly come from the large-angle reionisation
   signature~\citep{zaldarriaga97b} and
   $r=0.05$ may be possible if foregrounds allow. \planck\, will 
   be sensitive to the multipole range $2 < \ell <3000$ and is
   scheduled to launch in 2009. The HFI has no active or fast signal modulation
   (i.e.\ other than scanning).

  \vspace{3mm}\item{{\cmbpolbpol}:}
  Design studies have been conducted
  for satellite mission(s) dedicated to measuring primordial
  $B$-modes comprising $\sim 2000$ detectors with the ability to measure
  $0.001 < r < 0.01$ if foregrounds allow it~\citep{bock08,deBernardis08}.
  The timescale for launch of any selected mission is likely beyond 2020.
\end{itemize}

\section{Simulations}
\label{sec:sims}

\begin{figure}
  \centering
  \resizebox{0.48\textwidth}{!}{  
    \rotatebox{-90}{\includegraphics{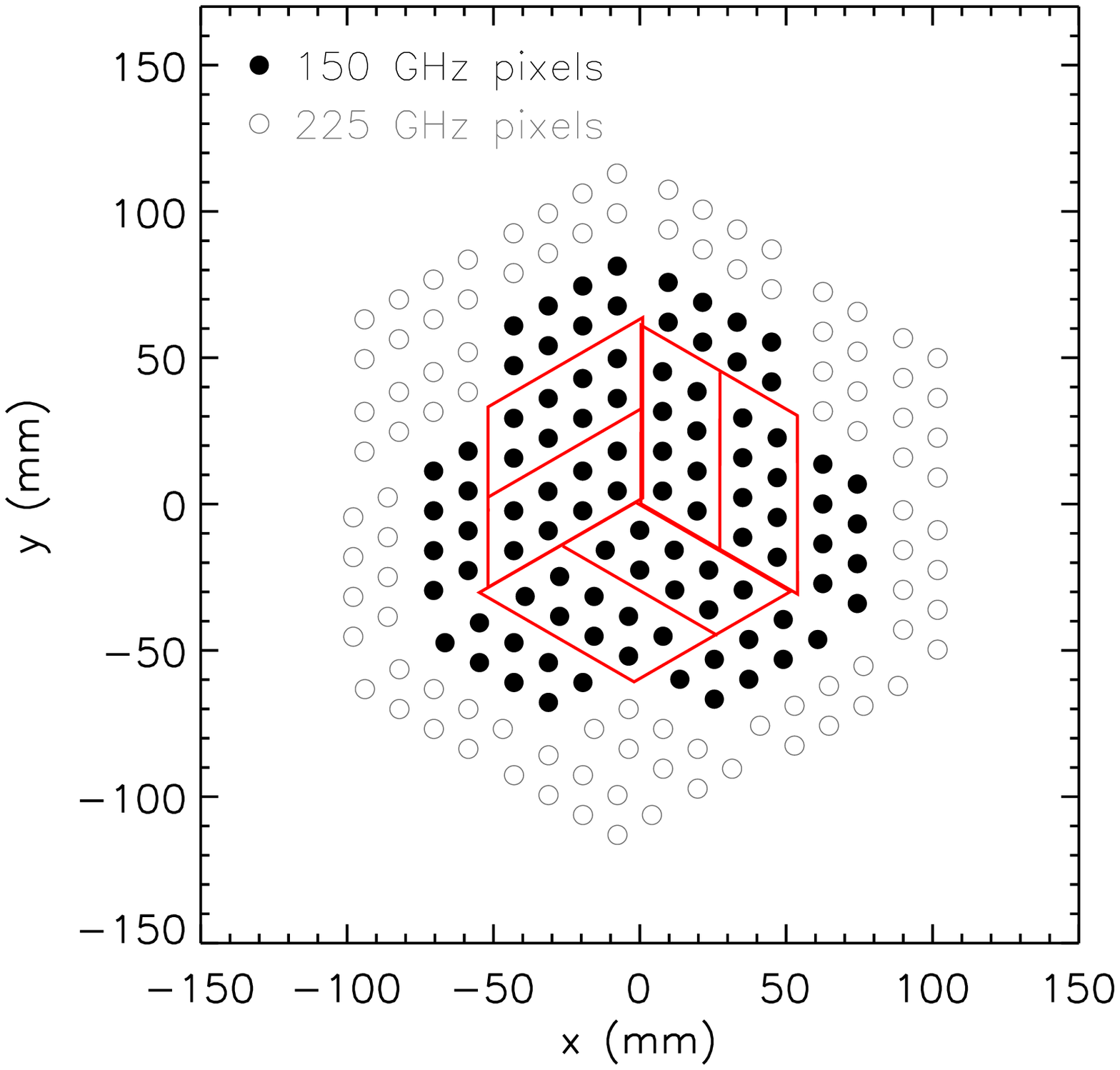}}}
  \caption{Layout of detectors on the \clover\, high frequency focal
  plane. Each point indicates a pixel comprising two TES detectors
  sensitive to orthogonal linear polarizations. The polarization
  sensitivity directions of detectors within each block are along, and
  at right angles to, the major axis of the block. The outlined
  150~GHz detector blocks at the centre of the array are used in the
  simulations described here. The field of view of the entire array is
  $\sim 5 \, {\rm deg}^2$.}
  \label{fig:hf_focal_plane}
\end{figure}

The \clover\, experiment will consist of two telescopes -- a low frequency
instrument with a focal plane consisting of 192
single-polarization 97~GHz TES detectors, and a high frequency instrument with
a combined focal plane of 150 and 225~GHz detectors (192 of
each). Note that we have not included foreground contamination in our
simulations, so for this analysis, we consider only the 150~GHz
detector complement -- the corresponding reduction in sensitivity will
approximate the effect of using the multi-frequency observations to
remove foregrounds.  Figure~\ref{fig:hf_focal_plane} shows the
arrangement of the 150 and 225~GHz detectors on the high frequency
focal plane. The detectors are arranged in detector blocks consisting
of eight pixels each. Each pixel consists of two TES detectors which are
sensitive to orthogonal linear polarizations. The polarization
sensitivity of the eight detector pairs within a block are along, and at
right angles to, the major axis of their parent block. The detector
complement, both at 150 and 225~GHz, therefore consists of three
`flavours' of pixels with different polarization sensitivity
directions. The 97~GHz focal plane (not shown) has a similar mix of
detector orientations.

\subsection{Simulation parameters}
\label{sec:sims_params}
Because simulating the full \clover\, experiment is computationally
demanding (a single simulation of a two-year campaign
at the full \clover\, data rate would require $\sim 10^4$ CPU-hours),
we have scaled some of the simulation parameters in order to make our
analysis feasible.
\begin{enumerate}
\item We simulate only half the 150~GHz detector complement and have
  scaled the noise accordingly. We have verified with a restricted
  number of simulations using all the 150~GHz detectors that the
  marginally more even coverage obtained across our field using all detectors
  has little or no impact on our results or
  conclusions. The 150~GHz detector blocks used in our simulations are
  indicated in Fig.~\ref{fig:hf_focal_plane} and include all three
  possible orientations of pixels on the focal plane.
\item The \clover\, detectors will have response times of $\sim 200\,
  \mu$s and so the data will be sampled at $\sim 1$~kHz in order to
  sample the detector response function adequately. Simulating at this
  rate is prohibitive so we simulate at a reduced data rate of
  $100$~Hz. For the \clover\, beam size (FWHM = 5.5 arcmin at 150~GHz) and
  our chosen scan speed ($0.25^{\circ}$/s), this data rate is still fast
  enough to sample the sky signal adequately.
\item \clover\, will observe four widely separated fields on the sky,
  each covering an area of $\sim 300 \, {\rm deg}^2$, over the course
  of two years. Two of the fields are in the southern sky and two
lie along the equator. For our analysis, we observe each of the four fields
  for a single night only and, again, we have scaled the noise levels to those
  appropriate for the full two-year observing campaign.
\end{enumerate}

\subsection{Observing strategy}
\label{sec:obs_strategy}
\begin{figure*}
  \centering
  \resizebox{0.80\textwidth}{!}{  
    \rotatebox{-90}{\includegraphics{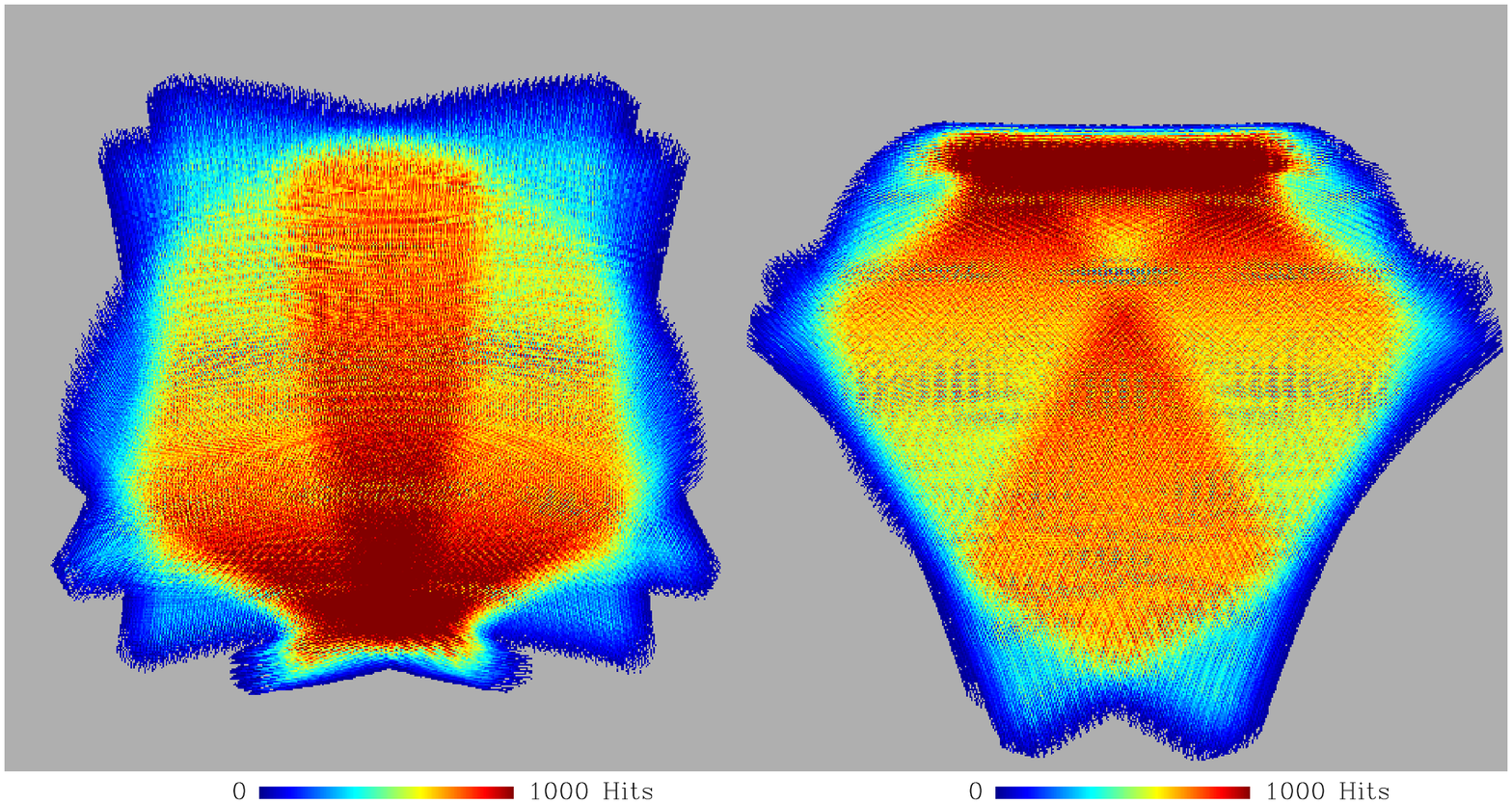}}}
  \caption{Hit-maps for one of the southern fields (left; RA
  09:30 hrs , Dec -40.00$^{\circ}$) and one of the equatorial fields
  (right; RA 04:00 hrs, Dec 0.00$^{\circ}$) for a single day's
  observation with half the 150~GHz detector complement. The central
  part of the fields (shown in yellow and red) are roughly $20^{\circ}$ in
  diameter. These maps have been constructed using a \healpix\, resolution of
  $N_{\rm side} = 1024$ corresponding to a pixel size $\sim$3.4 arcmin.}
  \label{fig:hitmaps}
\end{figure*}

Although optimisation of the observing strategy is not the focus of
this work, a number of possible strategies have been investigated by
the \clover\, team. For our analysis, we use the most favoured scan
strategy at the time of writing.  To minimise rapid variations in
atmospheric noise, the two telescopes will scan back and forth in
azimuth at constant elevation, allowing the field to rise through the
chosen observing elevation. Every few hours, the elevation angle of
the telescopes will be re-pointed to allow for
field-tracking. Although the precise details of the scan are likely to
change, the general characteristics of the scan and resulting field
coverage properties will remain approximately the same due to the
limitations imposed by constant-elevation scanning and observing from
Atacama. The \clover\, telescopes are designed with the capability of
scanning at up to $10^{\circ}$/s. However, for our analysis where we
have considered \clover\, operating with a rotating HWP (Section
\ref{sec:modulation}), we have chosen a relatively slow scan speed of
$0.25^{\circ}$/s in light of the HWP rotation frequency which we have
employed ($f_\lambda = 3$~Hz). Although the mode of operation of a HWP
on \clover\, is still under development, a continuously rotating HWP
is likely to be restricted to rotation frequencies of $f_\lambda <
5$~Hz due to mechanical constraints (with current cryogenic rotation
technologies, fast rotation, $\gsim 5$~Hz, could possibly result in
excessive heat generation). Figure~\ref{fig:hitmaps} shows the coverage maps for a
single day's observing on one of the southern fields and on one of the
equatorial fields. The corresponding maps for the other two fields are
broadly similar. Note that, in the real experiment, we expect that
somewhat more uniform field coverage than that shown in
Fig.~\ref{fig:hitmaps} will be achievable by employing slightly
different scan patterns on different days.

\subsection{Signal simulations}
\label{sec:signal_sims}
We generate model $TT$, $EE$, $TE$ and $BB$ CMB power spectra using
\camb\, \citep{lewis00}. The input cosmology used consists
of the best-fit standard $\Lambda$CDM model to the 5-year \wmap\, data
set \citep{hinshaw09}, but with a tensor-to-scalar ratio of $r=0.026$,
chosen to match the \clover\, `target' value. Realisations of CMB
skies from these power spectra are then created using a modified
version of the \healpix\footnote{%
See http://healpix.jpl.nasa.gov} 
software \citep{gorski05}. Our simulations include weak gravitational lensing
but ignore its non-Gaussian aspects.  Using only Gaussian simulations
means that we slightly mis-estimate the covariance matrices of our
power spectrum estimates, particularly the $B$-mode
covariances~\citep{smith04,smithchallinor06}. For the \clover\, noise
levels, this is expected to have a negligible impact on the
significance level of the total $B$-mode signal.

As part of the simulation process, the input CMB signal is convolved with a
perfect Gaussian beam with FWHM $=5.5$ arcmin. Note that an important
class of systematic effects which we do not consider in this paper are
those caused by imperfect optics; see the discussion in
Section~\ref{sec:systematics}.

For our analysis of the simulated data sets (Section
\ref{sec:analysis}) we have chosen to reconstruct maps of the Stokes
parameters with a map resolution of $3.4$ arcmin (\healpix\, $N_{\rm
side} = 1024$). Note that this pixel size does not fully sample the
beam and it is likely that we will adopt $N_{\rm side} = 2048$ for the
analysis of real data.  In order to isolate the effects of the various
systematics we have considered, our simulated CMB skies have therefore
been created at this same resolution --- we can then be sure that any
bias found in the recovered CMB signals is due to the systematic
effect under consideration rather than due to a poor choice of map
resolution\footnote{%
We have verified that spurious $B$-modes
generated through the pixelisation are negligible for the \clover\,
noise levels.}.  
Note that our adopted procedure of simulating and map-making at
identical resolutions, although useful for the specific aims of this
paper, is not a true representation of a CMB observation. For real
observations, pixelisation of the CMB maps will introduce a bias to
the measured signal on scales comparable to the pixel size adopted.
 
Using the pointing registers as provided by the scan strategy and
after applying the appropriate focal plane offsets for each detector,
we create simulated time-streams according to 
\be 
d_i = \left[ T(\theta) + Q(\theta) \cos(2\phi_i) + U(\theta) \sin(2\phi_i) \right]/2,
\label{eqn:signal_timestream}
\ee 
where $\theta$ denotes the pointing and $T, Q$ and $U$ are the sky
signals as interpolated from the
input CMB sky map. The polarization angle, $\phi_i$ is, in general, a
combination of the polarization sensitivity direction of each detector,
any rotation
of the telescope around its boresight, the direction of travel of the
telescope in RA--Dec space and the orientation of the half-wave plate,
if present.

\subsection{Noise simulations}
\label{sec:noise_sims}
The \clover\, data will be subject to several different noise
sources. Firstly, photon loading from the telescope, the atmosphere
and the CMB itself will subject the data to uncorrelated random
Gaussian noise. Secondly, the TES detectors used in \clover\, are
subject to their own sources of noise which will possibly
include low-frequency $1/f$ behaviour and correlations between
detectors. Thirdly, the atmosphere also has a very strong $1/f$
component which will be heavily correlated across the detector
array. Fortunately, the $1/f$ component of the atmosphere is known to
be almost completely un-polarized and so can be removed from the
polarization analysis by combining data from multiple detectors (see
Section \ref{sec:differencing}).

The white-noise levels due to loading from the instrument, atmosphere
and CMB have been carefully modelled for the case of \clover\,
observations from Atacama. We will not present the details here, but
for realistic observing conditions and scanning elevations, we have
calculated the expected noise-equivalent temperature (NET) due to
photon noise alone to be $\approx 146 \, \mu {\rm K} \sqrt{\rm s}$. We
add this white noise component to our simulated signal time-streams
for each detector as
\be
d_i \rightarrow d_i +  \frac{\rm NET}{2} \sqrt{f_{\rm samp}} g_i,
\ee
where $f_{\rm samp}$ is the sampling frequency and $g_i$ is a Gaussian
random number with $\mu = 0$ and $\sigma = 1$. Note that the white-noise
level in the detector time streams is ${\rm NET} / 2$ since the
\clover\, detectors are half-power detectors (equation
\ref{eqn:signal_timestream}). 

Using instrument parameters appropriate for the \clover\, detectors, 
we use the small-signal TES model of \cite{irwin05} to create a
model noise power spectrum for the detector noise. This model includes
both a contribution from the super-conducting SQUIDs which will be used to
record the detector signals (e.g. \citealt{reintsema03}) and a
contribution from aliasing in the Multi Channel Electronics (MCE;
\citealt{battistelli08}) which will be used to read out the signals. 
For the instrument parameters we have chosen, the effective
NET of the detector noise in our simulations is approximately equal
to the total combined photon noise contribution from the atmosphere, the
instrument and the CMB. Note however that for the final instrument, it
is hoped that the detector NET can be reduced to half that of the
total photon noise, thus making \clover\, limited by irreducible
photon loading.
The \cite{irwin05} small-signal TES model does not 
include a $1/f$ component to the detector noise so in order
to investigate the impact of modulation on possible low-frequency
detector noise, we add a heuristically chosen $1/f$ component to the
detector noise model with knee frequencies in the range, $0.01 <
f_{\rm knee} < 0.1$~Hz. 
The MCE system which will be used to read out the \clover\, data
should have low cross-talk between different channels. However,
correlations will be present at some level and so we include
10 per cent correlations between all of our simulated detector noise
time-streams. Generally, to simulate stationary noise that is correlated in time and
across the $N_{\rm det}$ detectors we proceed as follows.
Let the noise cross-power spectrum
between detector $d$ and $d'$ be $P_{d,d'}(f)$. 
Taking the Cholesky decomposition of this matrix at each frequency,
$L_{d,d'}(f)$, defined by
\be 
P_{d,d'}(f) = \sum_{d''} L_{d,d''}(f) L_{d',d''}(f), 
\ee
we apply $L$ to $N_{\rm det}$ independent, white-noise time streams
$g_{d}(f)$ in Fourier space,
\be
g_d(f) \rightarrow \sum_{d'} L_{d,d'}(f) g_{d'}(f) .
\ee
The resulting time-streams, transformed to
real space then possess the desired correlations between detectors. Here,
we assume that the correlations are independent of frequency so that
the noise cross-power spectrum takes the form
\be
P_{d,d'}(f) = C_{d,d'} P(f) ,
\ee
where the correlation matrix $C_{d,d'} = 1$ for $d = d'$ and $C_{d,d'} = 0.1$
otherwise. In practice, we use discrete Fourier transforms to synthesise
noise with periodic boundary conditions (and hence circulant time-time
correlations).

We use the same technique to simulate the correlated $1/f$ component
of the atmosphere. We have measured the noise properties of the atmosphere
from data from the \quad\, experiment \citep{hinderks09}. The 150~GHz
frequency channel of \quad\, is obviously well matched to the
\clover\, 150~GHz channel although \quad\, observed the CMB from the
South Pole rather than from Atacama. Although there are significant
differences between the properties of the atmosphere at the South Pole
and at Atacama (e.g. \citealt{bussmann05}), the \quad\, observations
still represent the best estimate of the $1/f$ noise properties of
the atmosphere available at present. A rough fit of the \quad\, data to the model, 
\be
P(f) = {\rm NET^2} \left[1 + \left( \frac{f_{\rm knee}}{f} \right)^\alpha \right],
\label{eqn:pk_atms}
\ee
yields a knee frequency, $f_{\rm knee} = 0.45$~Hz and spectral index,
$\alpha = 2.5$. Using this model power spectrum we simulated $1/f$
atmospheric noise correlated across the array in exactly the same way
as was used for the detector noise. Fortunately, the $1/f$ component
in the atmosphere is almost completely un-polarized.
If there were no instrumental polarization, detectors
within the same pixel (which always look in
the same direction) would therefore be completely correlated
with one another and detectors from different pixels would also be
heavily correlated. For the correlated atmosphere, we therefore use a
correlation matrix given by $C_{d,d'} = 1.0$ for $|d-d'| \le 1$
(i.e. for detector pairs) and $C_{d,d'} = 0.5$ otherwise. In the
following sections, as one of the systematics we have investigated, we 
relax the assumption that the atmosphere is un-polarized.

\begin{figure}
  \centering
  \resizebox{0.47\textwidth}{!}{  
    \rotatebox{-90}{\includegraphics{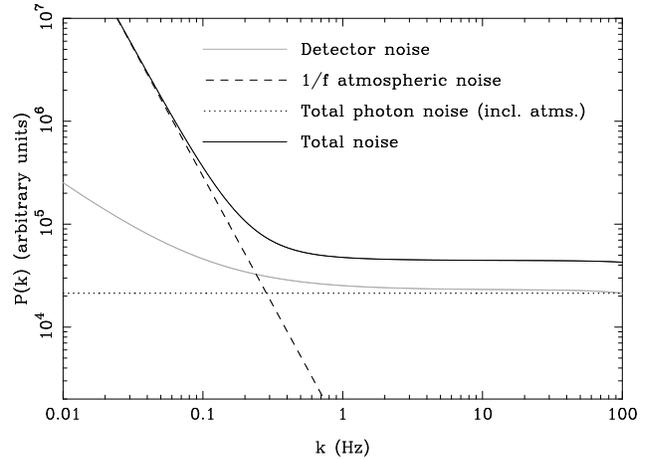}}}
  \caption{Frequency space comparison between the different noise
    sources in the simulations. The grey line shows the detector noise
    power spectrum (here with a $1/f$ component with knee frequency,
    $f_{\rm knee} = 0.1$~Hz). The correlated component to the
    atmosphere is shown as the dashed line and the total photon noise
    (including atmospheric loading) is shown as the dotted line. For
    the simulation parameters we have adopted, the temperature sky-signal from
    multipoles $20 < \ell < 2000$ appears in the time-stream in the
    frequency range $0.01 < f < 1$~Hz. In the absence of fast modulation,
    the polarized sky signal also appears in this frequency range
    whereas in our simulations including fast modulation, the polarized
    sky signal appears in a narrow band centred on 12~Hz.}
  \label{fig:noise_compare}
\end{figure}

\begin{figure}
  \centering
  \resizebox{0.48\textwidth}{!}{  
    \rotatebox{-90}{\includegraphics{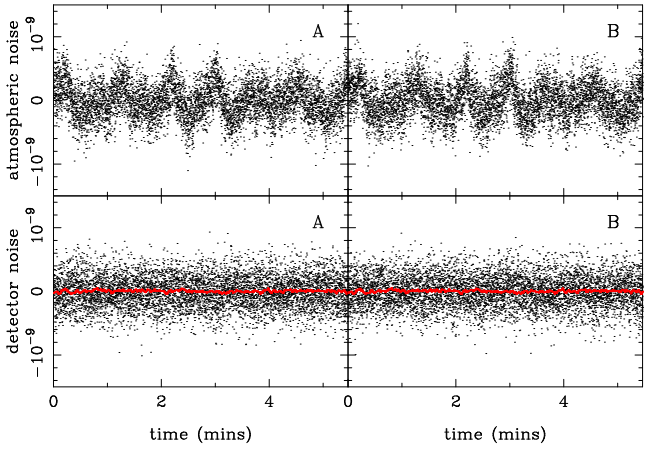}}}
  \caption{A five-minute sample of simulated noise time-stream for the two
  detectors within a pixel (denoted ``A'' and ``B'') for both the
  atmospheric noise simulations (\emph{top}) and for the detector
  noise simulations (\emph{bottom}). The $1/f$ component in the
  atmospheric noise time-streams is 100 per cent correlated between the
  A and B detectors and can be removed entirely by
  differencing. In contrast the $1/f$ component in the detector noise
  (which is much weaker and not noticeable on this plot)
  is only weakly correlated between A and B and is not removed by
  differencing. For comparison, the signal-only time-streams are
  also plotted in the lower panels as the red curves.}
  \label{fig:tod_noise}
\end{figure}

Figure~\ref{fig:noise_compare} compares the photon, atmospheric $1/f$
and detector noise contributions to our simulated data in frequency
space. At low frequencies, the noise is completely dominated by the
atmospheric $1/f$ while the white-noise contributions from photon
loading (including the uncorrelated component of the atmosphere) and
detector noise are approximately equal. Note that for observations
without active modulation, and for the scan speed and observing
elevations which we have adopted, the ``science band'' for the
multipole range $20 < \ell < 2000$ corresponds roughly to $0.01 < f <
1$~Hz in time-stream frequency. In contrast, for our simulations which
include a continuously rotating HWP, the temperature signal remains within
the $0.01 < f < 1$~Hz frequency range but the polarized sky signal is
moved to a narrow band centred on $\sim$12~Hz, well away from both the
detector and atmospheric $1/f$ noise components (see
Section~\ref{sec:modulation} and Fig.~\ref{fig:mod_frequency}). Note also that although the
atmospheric $1/f$ dominates the detector $1/f$ at low frequency, the
atmosphere is heavily correlated across detectors and can therefore be
removed by combining detectors (e.g. differencing detectors within a
pixel) but this is not true for the detector noise which is only
weakly correlated between detectors. We demonstrate this in
Fig.~\ref{fig:tod_noise} where we plot a five-minute sample of simulated
atmospheric and detector noise for the two constituent detectors
within a pixel. Including the atmospheric $1/f$ component, the
effective total NET per detector measured from our simulated
time-streams is $293 \, \mu {\rm K} \sqrt{\rm sec}$ whilst excluding
atmospheric $1/f$, we measure $210 \, \mu {\rm K} \sqrt{\rm s}$.


\subsection{Detector response}
\label{sec:detector_response}
The \cite{irwin05} TES small-signal model mentioned above also
provides us with an estimate of the detector response (the conversion
from incident power to resultant current in the detectors). In this
model, the power-to-current responsivity, $s_I(\omega)$, is given by
\be 
s_I(\omega) \propto \frac{1 - \tau_+/\tau_I}{1 + i\omega\tau_+}
\frac{1 - \tau_-/\tau_I}{1 + i\omega\tau_-}, 
\label{eqn:response}
\ee 
where $\omega = 2 \pi f$ is the angular frequency and $\tau_+$ and
$\tau_-$ are the ``rise time'' and ``fall time'' (relaxation to steady
state) after a delta-function temperature impulse. Here, $\tau_I$ is
the current-biased thermal time constant. The impulse-response in
the time domain is
\be
s_I(t) \propto \frac{e^{-t/\tau_+} - e^{-t/\tau_-}}{\tau_+ - \tau_-}
\Theta(t) ,
\ee
where $\Theta(t)$ is the Heaviside step function.
Note that the constant of
proportionality in equation~(\ref{eqn:response}) (which is also
predicted by the model) is essentially the calibration (gain) of the
detectors. The simulated time-streams of
equation~(\ref{eqn:signal_timestream}) are converted to detector
time-streams through convolution with this response
function. Note that the photon noise and correlated atmospheric noise
are added to the time-stream before convolution (and so are also
convolved with the response function) while the detector noise is
added directly to the convolved time-streams. The \clover\, detectors
are designed to be extremely fast with time-constants $\tau_\pm < 1 \,
{\rm ms}$. For our simulations, we have used time-constants
predicted by the small-signal TES model for \clover\, instrument
parameters of $\tau_+ = 300 \, \mu{\rm s}$ and $\tau_- = 322 \, \mu
{\rm s}$. 

Note that for our chosen scan speed of $0.25^{\circ}$/s, the effect of
the response function on the signal component in the simulations is
small in the absence of fast modulation -- the \clover\,
detectors are so fast that signal attenuation
and phase differences introduced by convolution with the detector
response only become important at high frequencies, beyond the
frequency range of the sky signals. For our simulations without
fast modulation, sky signals from multipoles $\ell \sim 2000$ will
appear in the time-stream at $\sim$ 1~Hz where the amplitude of the
normalised response function is effectively unity and the associated
phase change is $-0.2^{\circ}$. For our simulations
including fast modulation, it becomes more important to correct the
time-stream data for the detector response. In this case, the polarized sky signals (from all
multipoles) appear within a narrow band centred on $12$~Hz in the
time-stream. Here, the amplitude of the response function is still very
close to unity but the phase change has grown to $-2.7^{\circ}$. We
include a deconvolution step in the analysis of all of our simulated
data to correct for this effect. Finally, we note that for the $\sim 10$
per cent errors in the time-constants which we have considered (see
Section~\ref{sec:systematics}), the resulting mis-estimation of the
polarization signal will again be small, even for the case of
fast modulation.

In principle we should also include the effect of sample integration: each
discrete observation is an integral of a continuous signal over the sample period.
Of course, in the case where down-sampled data is simulated
the sample integration should include the effect of the sample
averaging. For integration over a down-sampled period $\Delta$, there is an additional
phase-preserving filtering by $\mathrm{sinc}(\omega \Delta/2)$ where
$\omega \equiv 2\pi f$. For our scan
parameters, the filter is negligible (i.e.\ unity) for unmodulated data
while for modulated data the filter can be approximated at the frequency
$4 \omega_\lambda$, where $\omega_\lambda$ is the angular rotation frequency
of the HWP. This acts like a small decrease in the polarization efficiency,
with the discretised signal at the detector being
\begin{eqnarray}
d_i &\approx & \frac{1}{2}\big\{T(\theta) +
\mathrm{sinc}(2\omega_\lambda \Delta) \nonumber \\
&&\mbox{} \times \left[Q(\theta)\cos(2\phi_i)
+ U(\theta) \sin(2\phi_i)\right]\big\} .
\end{eqnarray}
For $\omega_\lambda = 2\pi \times 3\,\mathrm{Hz}$ and $\Delta =
10\,\mathrm{ms}$, the effective polarization efficiency is $0.98$ and
this has the effect of raising the noise level in the polarization maps by
2 per cent. We do not include this small effect in our simulations, but could
easily do so.

\subsection{Systematic effects}
\label{sec:systematics}
In the analysis that follows, we will investigate the impact of
several systematic effects on the ability of a \clover-like
experiment to recover an input $B$-mode signal. For a reference, we
use a suite of simulations which contain no systematics. This ideal simulation
contains the input signal, photon noise, $1/f$ atmospheric noise
correlated across the array (but un-polarized) and TES
detector noise with no additional correlated $1/f$
component. Additionally, for our reference simulation all pointing
registers and detector polarization sensitivity angles use the nominal
values and the signal is convolved with the detector response function
using the nominal time constants.

We then perform additional sets of simulations with the following
systematic effects included in isolation:
\begin{itemize}
\item $1/f$ detector noise. We have considered an additional
  correlated $1/f$ component to the detector noise with $1/f$ knee 
  frequencies of $0.1$, $0.05$ and $0.01$~Hz.
\item Polarized atmosphere. In addition to the un-polarized atmosphere
  present in the reference simulation, we consider a \emph{polarized}
  $1/f$ component in the atmosphere. To simulate polarized atmosphere,
  we proceed as described in Section~\ref{sec:noise_sims} but now
  we add correlated $1/f$ atmospheric noise to the $Q$ and $U$ sky signal
  time-streams such that equation~(\ref{eqn:signal_timestream})
  becomes 
  \begin{eqnarray}
  d_i &=& \frac{1}{2} \left[T + \left(Q + Q^{\rm atms}_i\right) \cos(2 \phi_i)
  \nonumber \right. \\
  && \mbox{} \phantom{xxxx} \left. 
  + \left(U + U^{\rm atms}_i\right) \sin(2 \phi_i)\right]. 
  \end{eqnarray}
  We take the $Q^{\rm atms}$ and $U^{\rm atms}$ atmospheric signals to
  have the same power spectrum as the common-mode atmosphere
  (equation~\ref{eqn:pk_atms}) but a factor ten smaller in magnitude. 
\item Detector gain errors. We consider three types of gain errors:
  (i) random errors in the gain that are constant in time, uncorrelated
  between detectors and have a 1 per cent RMS; (ii) gain drifts
  in each detector corresponding to a 1 per cent drift over the course of a
  two-hour observation -- the start and end
  gains for each detector are randomly distributed about the nominal
  gain value with an RMS of 1 per cent; (iii) systematic A/B gain
  mis-matches (1 per cent mis-match) between the two detectors within each
  pixel. For this latter systematic, we have applied a constant 1 per
  cent A/B mis-match to all pixels on the focal plane but the direction
  of the mis-match (that is, whether the gain of A is greater or smaller
  than B) is chosen randomly.
 \item Mis-estimated polarization sensitivity directions. Random
 errors uncorrelated between detectors (including those with the
 same feedhorn) with RMS 0.5$^{\circ}$ and which are constant in time,
 and a systematic mis-estimation of the
 instrument polarization coordinate reference system by 0.5$^{\circ}$ are
 considered.
\item Mis-estimated half-wave plate angles. For the case where we
  consider an experiment which includes polarization modulation with a
  half-wave plate (see Section \ref{sec:modulation}), we also
  consider random errors (with RMS 0.5$^{\circ}$) in the recorded HWP angle 
  which we apply to each 100~Hz time-sample. In addition, we consider
  a 0.5$^{\circ}$ systematic offset in the half-wave plate angle measurements.
\item Mis-estimated time-constants. The analysis that follows
  includes a deconvolution step to undo the response function of the
  detectors and return the deconvolved sky signal. In all cases, we
  use the nominal time-constant values of $\tau_+ = 300 \, \mu{\rm s}$
  and $\tau_- = 322 \, \mu {\rm s}$ to perform the deconvolution. To
  simulate the effect of mis-estimated time-constants, we introduce
  both a random scatter (with RMS = 10 per cent across detectors) and a
  systematic offset ($\tau_\pm$ identically offset by 10 per cent for all
  detectors) in the time-constants when creating the simulated data.
\item Pointing errors. We simulate the effects of both a random jitter and a
  slow wander in the overall pointing of the telescope by introducing
  a random scatter uncorrelated between time samples (with RMS 30 arcsec)
  and an overall drift in the
  pointing (1 arcmin drift from true pointing over the course of a
  two-hour observation) when creating the simulated time-stream. Once
  again, the simulated data is subsequently analysed assuming perfect
  pointing registers.
\item Differential transmittance in the HWP. As a simple example of a
  HWP-induced systematic, we have considered a differential
  transmittance of the two linear polarizations by the
  HWP. Preliminary measurements of the \clover\, HWPs suggest the level
  of differential transmittance will be in the region of
  $1\!\!\!\relbar\!\!\!2$ per cent. For
  this work, we consider a 2 per cent differential transmittance in the HWP. 
\end{itemize}

The range of systematic effects we include is not exhaustive. In
particular, we ignore effects in the HWP, when present, except for a
mis-estimation of the rotation angle and a differential transmittance
of the two linear polarizations. In addition, we ignore all optical
effects. In practice, there are many possible HWP-related systematic
effects which we have not yet considered. In general, a
thorough analysis of HWP-related systematics require detailed physical
optics modelling which is beyond the scope of our current analysis. We
therefore leave a detailed investigation of HWP-related systematics to future
work and simply urge the reader to bear in mind that where our
analysis has included a HWP, we have, in most cases, assumed a perfect one.  For other
optical effects, we note that \citet{odea07} have already investigated
some relevant effects using analytic and numerical techniques and we
are currently adapting their flat-sky numerical analysis to work with
the full-sky simulations described here. Our conclusions on the
ability of modulation to mitigate systematic effects associated with
imperfect optics, which will be based on detailed physical optics
simulations of the \clover\, beams (Johnson et al., in prep), will be
presented in a future paper.  Note that there are important
instrument-specific issues to consider in such a study to do with
where the modulation is performed in the instrument. In \clover, the
modulation will be provided by a HWP between the horns and mirrors and
this may lead to a difference in the relative rotation of the field
directions on the sky and the beam shapes as the HWP rotates compared
to a set-up, as in \spider~\citep{crill08}, where the HWP is after
(thinking in emission) the beam-defining elements.

For a \clover-like receiver, consisting of a HWP followed by a polarization
analyser (e.g.\ an orthomode transducer) and detectors, we include
essentially all
relevant systematic effects introduced \emph{by the receiver}.
To see this, note that
the most general Jones matrix describing propagation of the two linear
(i.e.\ $x$ and $y$) polarization states through the polarization analyser
is~\citep{odea07}
\be
\mathbfss{J} = \left(
\begin{array}{cc}
1+g_1 & \epsilon_1 \\
-\epsilon_2 & (1+g_2)e^{i\alpha}
\end{array}
\right) ,
\ee
where $g_1$, $g_2$ and $\alpha$ are small real parameters and
$\epsilon_1$ and $\epsilon_2$ are small and complex-valued. The detector
outputs are proportional to the power in the
$x$ and $y$-components of the transmitted field (after convolution
with the detector response function).
To first-order in small
parameters, only $g_1$, $g_2$ and the real parts of $\epsilon_1$ and
$\epsilon_2$ enter the detected power. In this limit, the perturbed Jones
matrix is therefore equivalent in terms of the detected power to
\be
\mathbfss{J} \sim \left(
\begin{array}{cc}
1+g_1 & 0 \\
0 & 1+g_2
\end{array}
\right)
\left(
\begin{array}{cc}
\cos \alpha_1 & \sin \alpha_1 \\
-\sin\alpha_2 & \cos\alpha_2
\end{array}
\right)
\ee
where the small angles $\alpha_1 \approx \Re\epsilon_1$ and $\alpha_2 \approx
-\Re\epsilon_2$ denote the perturbations in the polarization-sensitivity
directions introduced above, and $g_1$ and $g_2$ are the gain errors.
Note that instrument polarization (i.e.\ leakage from $T$ to detected 
$Q$ or $U$) is only generated in the receiver through mismatches in the gain at first 
order, but also through $|\epsilon_1|^2$ and
$|\epsilon_2|^2$ in an exact calculation. The latter effect is not
present in the simplified description in terms of offsets in the
polarization-sensitivity directions. Note also that if we
difference the outputs of
the two detectors in the same pixel, in the presence of perturbations
$\alpha_1$ and $\alpha_2$ to the polarization sensitivity directions we
find
\begin{eqnarray}
d_1 - d_2 &=& \cos(\alpha_1-\alpha_2)\left(
Q\cos[2(\phi-\bar{\alpha})] \nonumber \right. \\
&&\mbox{} \phantom{xxxxxxxxxxxx} \left. + U \sin[ 2(\phi-\bar{\alpha})] \right) ,
\end{eqnarray}
where $\bar{\alpha} = (\alpha_1 + \alpha_2)/2$. This is equivalent
to a common rotation of the pair by $\bar{\alpha}$ and a decrease in the
polarization efficiency to $\cos(\alpha_1-\alpha_2)$.

\subsection{Polarization modulation}
\label{sec:modulation}

For our reference simulation, and for each of the systematic effects
listed above, we simulate the experiment using three different
strategies for modulating the polarization signal. Firstly, we
consider the case where no explicit modulation of the polarization
signal is performed -- in this case, the only modulation achieved is
via telescope scanning and the relatively small amount of sky rotation
provided by the current \clover\, observing strategies. In addition,
we also consider the addition of a half-wave plate, either continuously
rotating or ``stepped'', placed in front of the focal plane. A
half-wave plate modulates the polarization signal such that the
output of a single detector (in the detector's local polarization
coordinate frame) is 
\be d_i = \frac{1}{2}\left[ T + Q \cos ( 4\phi_i )
+ U \sin ( 4\phi_i ) \right],  
\label{eqn:pol_mod}
\ee 
where, here, $\phi_i$ is the angle between the detector's local
polarization frame and the principal axes of the wave plate.

\begin{figure}
  \centering
  \resizebox{0.47\textwidth}{!}{  
    \rotatebox{-90}{\includegraphics{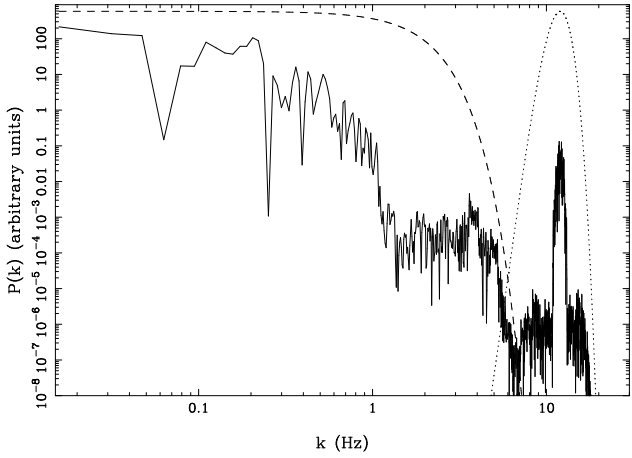}}}
  \caption{Frequency-space representation of polarization modulation
  with a continuously rotating HWP. The plotted power spectrum is that
  for a single azimuth scan from one of our signal-only simulations
  with the HWP continuously rotating at $f_\lambda = 3$~Hz. The power
  in the frequency range $0.01 < k < 1$~Hz is the unmodulated
  temperature signal from sky multipoles in the range $20 < \ell <
  2000$. The power spike at $4 f_\lambda = 12$~Hz is the modulated
  polarized signal. The dashed (dotted) line shows the expected
  temperature (polarization) signal band (with arbitrary
  normalisation) appropriate for the scan speed, modulation frequency
  and beam size we have used. The residual power between $1$ and
  $6$~Hz and on either side of the polarization power spike is due to
  pixelisation effects.}
  \label{fig:mod_frequency}
\end{figure}

For a continuously rotating HWP, the polarized-sky signal is thus
modulated at $4 f_\lambda$ where $f_\lambda$ is the rotation frequency
of the HWP. As well as allowing all three Stokes parameters to be
measured from a single detector, modulation with a continuously
rotating HWP (which we term ``fast'' modulation in this paper) moves
the polarization sky-signal to higher frequency and thus away from any
low-frequency $1/f$ detector noise that may be
present; see Fig.~\ref{fig:mod_frequency}.
(Note that the temperature signal is not
modulated and one needs to rely on telescope scanning and analysis
techniques to mitigate $1/f$ noise in $T$.) This ability to mitigate
the effect of $1/f$ noise on the polarization signal is the prime
motivation for including a continuously rotating HWP in a CMB
polarization experiment\footnote{%
The ability to measure all three Stokes
parameters from a single detector has also been suggested as
motivation for including a modulation scheme in CMB polarization
experiments. However, we will argue later in
Section~\ref{sec:differencing} that, at least for ground-based
experiments, an analysis based on extracting all three Stokes
parameters from individual detectors in isolation using a real-space
demodulation technique may be a poor choice.}
Systematic effects that generate an apparent polarization signal that is
not modulated at $4f_\lambda$ can also be mitigated almost
completely with fast modulation.
Most notably, instrument polarization generated in the receiver will
not produce a spurious polarization signal in the recovered maps
unless the gain and
time-constant mismatches vary sufficiently rapidly ($\sim 4 f_\lambda$)
to move the \emph{scan}-modulated temperature leakage up into the
polarization signal band. 

As mentioned in the previous section, as an example of a HWP-induced
systematic, we have considered the case of a 2 per cent differential
transmittance by the HWP of the two incoming linear polarizations. We
model this effect using a non-ideal Jones matrix for the HWP of the
form, 
\be
\mathbfss{J} = \left(
\begin{array}{cc}
1 & 0 \\
0 & -(1+\delta)
\end{array}
\right) ,
\label{eqn:diff_trans_jones}
\ee
where $\delta$ describes the level of differential
transmittance. Propagating through to detected power, for the
difference in output of the two detectors within a pixel, we find
\ba
d_1 - d_2 \, &=& \left[1 + \delta + \frac{\delta^2}{4} \right]
(Q\cos(4\phi) + U\sin(4\phi) ) \nonumber \\
             &-& \left[\delta + \frac{\delta^2}{2}\right] I\cos(2\phi)
              + \frac{\delta^2}{4} Q.
\label{eqn:diff_trans_power}
\ea
Note that, in this expression, both the HWP angle, $\phi$, and the
Stokes parameters are defined in the pixel basis. The first term in
equation (\ref{eqn:diff_trans_power}) is the ideal
detector-differenced signal but mis-calibrated by a factor, $\delta +
\delta^2/4$. For reasonable values of $\delta$, this mis-calibration
is small ($\lsim 2$ per cent) and, in any case, is easily dealt with during 
a likelihood analysis of the power spectra. The
potential problem term is the middle term which contains the
total intensity signal modulated at $2 f_\lambda$. Note that there
will be contributions to this term from the CMB monopole, dipole and
the atmosphere, which, for our simulations, we have taken to be 5 per cent
emissive. Even for small values of $\delta$ therefore, these
HWP-synchronous signals will completely dominate the raw detector data
and need to be removed from the data prior to the map-making step.

With a HWP operating in ``stepped'' mode where the angle of the HWP is
changed at regular time intervals (e.g. at the end of each scan), the
gains are less clear. The polarization sky signal is not shifted to
higher frequencies so $1/f$ detector noise can only be dealt with by
fast scanning. What stepping the waveplate can potentially do is to increase
the range of polarization sensitivity directions with which a given pair of detectors
samples any pixel on the sky. This has two important effects: (i) 
it reduces the correlations between the errors in the reconstructed $Q$ and $U$
Stokes parameters in each sky pixel; and (ii) it can mitigate somewhat those
systematic effects that do not transform like a true polarization under 
rotation of the waveplate. Of course, one of the strongest motivations
for stepped, slow modulation is the avoidance of systematic effects
associated with the continuous rotation of the HWP. If these effects are
sufficiently well understood, then the resulting spurious signals can
be rejected during analysis. However, if they are not well understood,
a stepped HWP, while not as effective in mitigating systematics, may
well be the preferred option.

Note that for a perfect optical system, rotation of the waveplate is
equivalent to rotation (by twice the angle) of the instrument.
However, this need not hold with imperfect optics. For example,
suppose the beam patterns for the two polarizations of a given
feedhorn are purely co-polar (i.e.\ the polarization sensitivity
directions are ``constant'' across the beams and orthogonal), but the
beam shapes are orthogonal ellipses. This set-up generates instrument
polarization with the result that a temperature distribution that is
locally quadrupolar on the sky will generate spurious polarization
that transforms like true sky polarization under rotation of the
instrument~\citep{hu03,odea07}. However, for an optical arrangement
like that in \spider, where the HWP is after (in emission) the
beam-defining optics, as the HWP rotates the polarization directions
rotate on the sky but the beam shapes remain fixed. The spurious
polarization from the mis-match of beam shapes is then constant as the
HWP rotates for any temperature distribution on the sky, and so the
quadrupolar temperature leakage can be reduced.

In our analysis, in addition to simulations without explicit
modulation we have also simulated an experiment with a HWP continuously
rotating at $3$~Hz (thus modulating the polarization signal at
$12$~Hz) and an experiment where a HWP is stepped (by 20$^{\circ}$) at the end
of each azimuth scan (for the scan strategy and scan speed we are
using, this corresponds to stepping the HWP roughly every $\sim$ 90
s).

We end this section with a comment on the ability of a continuously
rotating HWP to mitigate un-polarized $1/f$ atmospheric noise. The polarization
signal band is still, of course, moved to higher frequency and thus
away from the $1/f$ noise but the atmospheric $1/f$ noise is so strong
that extremely rapid HWP rotation would be required to move the
polarization band far enough into the tail of the $1/f$ spectrum.
Such rapid rotation is not an option in practice as it would introduce its
own systematic effects (e.g. excessive heat generation). This is the basis of our argument
mentioned above that extracting all three Stokes parameters from a
single detector may be a poor choice of analysis technique. However,
since the $1/f$ atmospheric noise is un-polarized, it can be removed
\emph{completely} by combining data from multiple detectors. We
revisit this issue again with simulations in
Section~\ref{sec:differencing}. Finally, we note that if the atmosphere
does contain a polarized $1/f$ component, then we expect that this
will not be mitigated by modulation -- the polarized atmosphere would
be modulated in the same way as the sky signal and would
shift up in frequency accordingly.

\section{Analysis of simulated data}
\label{sec:analysis}
For our reference simulation, and for each of the systematic effects
and modulation strategies described in the previous section, we have
created a suite of 50 signal-only, noise-only and signal-plus-noise
simulated datasets. Our analysis of the signal-only data will be used
to investigate potential biases caused by the systematics while our
signal-plus-noise realisations are used together with the noise-only
simulations to investigate any degradation of the sensitivity of the
experiment due to the presence of the systematic effects.

Our analysis of each dataset consists of processing the data through
the stages of deconvolution for the bolometer response function,
polarization demodulation and map-making, and finally estimation of the
$E$- and $B$-mode power spectra. For any given single realisation these
processes are performed separately for each of the four observed
\clover\, fields -- that is, we make maps and measure power spectra for
each field separately. Since our fields are widely separated on the
sky, we can treat them as independent and combine the power spectra
measured from each using a simple weighted average to produce a single
set of $E$- and $B$-mode power spectra for each realisation of the
experiment. Note that even if our fields were not widely separated,
our procedure would still be unbiased (but sub-optimal) and 
correlations between the fields would be automatically taken
into account in our error analysis since, for any given realisation, 
the input signal for all four fields is taken from the same simulated
CMB sky. 

\subsection{Time-stream processing and map-making}
\label{sec:map-making}
We first deconvolve the time-stream data for the detector response in
frequency space using the response function of
equation~(\ref{eqn:response}) and using the nominal time-constants in
all cases. Once this is done, the data from detectors within each
pixel are differenced in order to remove both the CMB temperature signal and
the correlated $1/f$ component of the atmospheric noise. For the case
where the $1/f$ atmosphere is completely correlated between the two
detectors and in the absence of instrumental polarization and/or
calibration systematics, this process will remove the CMB $T$ signal and the
$1/f$ atmosphere completely. 

\begin{figure}
  \centering
  \resizebox{0.455\textwidth}{!}{  
    \rotatebox{-90}{\includegraphics{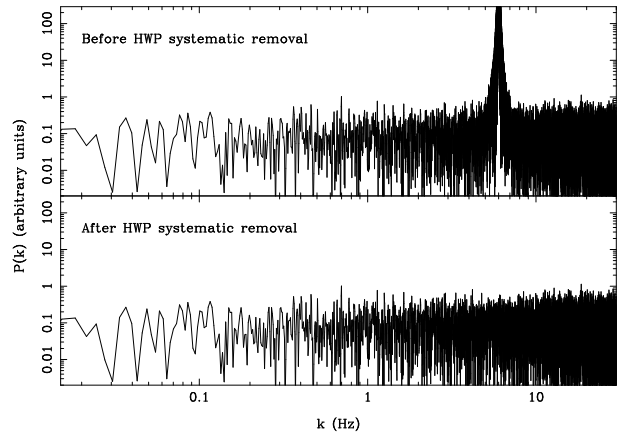}}}
  \caption{Power spectra of a six minute segment of time-stream
  data before (upper panel) and after (lower panel) fitting for and
  removing the HWP-synchronous signal. A 2 per cent differential
  transmittance in the HWP was used to create the simulated data. The
  resulting spurious signal appears at $2 f_\lambda = 6$~Hz in the upper
  panel and has a peak power of $\sim 10^{10}$ in the units
  plotted. No obvious residuals are apparent in the lower panel after
  applying our procedure for removing the contamination.}
  \label{fig:hwp_sys_removal}
\end{figure}

For the case where we have simulated the effect of a differential
transmittance in the HWP, a further time-stream processing step is
required at this point to fit for and remove the HWP-synchronous
signals from the time-stream. To do this, we have implemented a simple
iterative least-squares estimator to fit, in turn, for the amplitudes
of both a cosine and sine term at the second harmonic of the
HWP-rotation frequency, $f_\lambda$. For our simulations containing
both signal and noise, the accuracy with which we are able to remove
the HWP-synchronous signals is determined by the noise level in the
data. A demonstration of the performance of this procedure is given
in Fig.~\ref{fig:hwp_sys_removal} where we plot the power spectra of
six minutes of simulated time-stream data (containing both signal
and noise) before and after the removal of the HWP-synchronous signal.

After detector differencing, and the removal of the HWP-synchronous
signals if present, the resulting differenced time stream is then a
pure polarization signal:
\be d_i = Q \cos(2\phi_i) + U \sin(2\phi_i),
\label{eqn:diff_timestream}
\ee 
where again, the angle $\phi_i$ is, in the most general case, a
combination of detector orientation, sky crossing angle, telescope
boresight rotation and the orientation of the HWP if present.  In
order to construct maps of the $Q$ and $U$ Stokes parameters, these
quantities need to be decorrelated from the differenced time-stream
using multiple observations of the same region of sky taken with
different values of $\phi_i$. For an experiment which does not
continuously modulate the polarization signal, the $Q$ and $U$ signals
have to be demodulated as part of the map-making step. Note however
that for an experiment where the polarization signal is continuously
modulated, there are a number of alternative techniques to demodulate
$Q$ and $U$ at the time-stream level. In separate work, one of us has
compared the performance of a number of such demodulation schemes and
our results will be presented in a forthcoming paper (Brown, in prep.).
For the purposes of our current analysis
however, we have applied the same map-based demodulation scheme to all
three experiments which we have simulated. In this scheme, once the
time-streams from each detector pair have been differenced, $Q$ and
$U$ maps are constructed as 
\ba
\left( \begin{array}{c} Q \\ U \end{array} \right) &=& \left(
  \begin{array}{cc} 
    \lgl\cos^2(2\phi_i)\rgl & \lgl\cos(2\phi_i)\sin(2\phi_i)\rgl \\
    \lgl\cos(2\phi_i)\sin(2\phi_i)\rgl & \lgl\sin^2(2\phi_i)\rgl \\ 
\end{array} \right)^{-1} \nn \\
&&\mbox{} \times \left( \begin{array}{c}
    \lgl\cos(2\phi_i)d_i\rgl \\
    \lgl\sin(2\phi_i)d_i\rgl \\ \end{array} \right),
\label{eqn:qu_mapmaking}
\ea
where the angled brackets denote an average over all data falling within
each map pixel. For the work presented here, this averaging is
performed using the data from all detector pairs in one operation. 
One could alternatively make maps per detector pair which could then
be co-added later. For the case where the noise properties of each
detector pair are similar, the two approaches should be
equivalent. Note that the map-maker we use for all of our analyses is
a na\"{\i}ve one -- that is, we use simple binning to implement
equation~(\ref{eqn:qu_mapmaking}). There are, of course, more optimal
techniques available (e.g. \citealt{sutton09} and references therein)
which would out-perform a
na\"{\i}ve map-maker in the presence of non-white noise. However, for our
purposes, where we wish to investigate the impact of modulation in
isolation, it is more appropriate to apply the same na\"{\i}ve map-making
technique to all of our simulated data. We can then be sure that any
improvement we see in results from our simulations including
modulation are solely due to the modulation scheme employed. 

\subsection{Power spectrum estimation}
\label{subsec:power}
We measure $E$- and $B$-mode power spectra from each of our
reconstructed maps using the ``pure'' pseudo-$C_\ell$ method of
\cite{smith06}. We will not describe the method in detail here and refer the
interested reader to \cite{smith06} and \cite{smith07} for
further details. Here, we simply note that the ``pure''
pseudo-$C_\ell$ framework satisfies most of the requirements of a power
spectrum estimator for a mega-pixel CMB polarization experiment with
complicated noise properties targeted at constraining $B$-modes: it
is, just like normal pseudo-$C_\ell$, a fast estimator scaling as
$N_{\rm pix}^{3/2}$ where $N_{\rm pix}$ is the number of map pixels
(as opposed to a maximum likelihood estimator which scales as $N_{\rm
pix}^3$); it is a Monte-Carlo based estimator relying on simulations 
of the noise properties of the experiment to remove the noise bias and 
estimate band-power errors and covariances -- it is thus naturally
suited to experiments with complicated noise properties for which 
approximations to the noise cannot be made; and it is near-optimal in the
sense that it eliminates excess sample variance from $E \rightarrow B$ mixing
due to ambiguous modes which result from incomplete sky observations
\citep{lewis02, bunn02}, and which renders simple pseudo-$C_\ell$ techniques
unsuitable for small survey areas~\citep{challinor05}.

\subsubsection{Power spectrum weight functions}

With normal pseudo-$C_\ell$ estimators, one multiplies the data with a function
$W(\vnhat)$ that is chosen heuristically and apodizes the edge of the survey
(to reduce mode-coupling effects). For example, if one
is signal dominated, uniform weighting (plus apodization) is a reasonable
choice, whereas an inverse-variance weight is a good choice in the
noise-dominated regime. Similar reasoning applies for the
pure pseudo-$C_\ell$ technique, but here one weights the spherical
harmonic functions rather than the data themselves.
To see this, compare the
definition of the ordinary and pure pseudo harmonic $B$-modes:
\ba 
\widetilde a_{\ell m}^B = &-&\frac{i}{2}
\sqrt{\frac{(l-2)!}{(l+2)!}} \int d^2\vnhat \bigg[ \Pi_+(\vnhat)
W(\vnhat) \bar{\eth}\bar{\eth} Y_{\ell m}^*(\vnhat) \nonumber \\ 
&&\mbox{} - \Pi_-(\vnhat) W(\vnhat) \eth\eth Y_{\ell m}^*(\vnhat) \bigg]
\label{eq:almbdef} \\
\widetilde a_{\ell m}^{B\,{\rm pure}} = &-&\frac{i}{2}
\sqrt{\frac{(l-2)!}{(l+2)!}} \int d^2\vnhat \bigg[ \Pi_+(\vnhat)
\bar{\eth}\bar{\eth} \big( W(\vnhat) Y_{\ell m}^*(\vnhat) \big)
\nonumber \\ 
&&\mbox{} - \Pi_-(\vnhat) \eth\eth \big(
W(\vnhat) Y_{\ell m}^*(\vnhat) \big) \bigg]. 
\label{eq:almbpuredef} 
\ea
Here, $\Pi_\pm(\hat{n}) = (Q\pm
iU)(\vnhat)$ is the complex polarization and $\eth, \bar{\eth}$ are
the spin raising and lowering operators defined
in~\citet{zaldarriaga97}. If $W(\vnhat)$ is chosen so that it vanishes
along with its first derivative on the survey boundary, then the
$\widetilde a_{\ell m}^{B\,{\rm pure}}$ couple only to $B$-modes and
the excess sample variance due to $E$-$B$ mixing is eliminated.
The action of $\eth, \bar{\eth}$ on the spin spherical
harmonics is simply to convert between different spin-harmonics but
their action on a general weight function is non-trivial for
$W(\vnhat)$ defined on an irregular pixelisation such as \healpix\footnote{%
One possibility that we have yet to explore is performing the derivatives
directly in spherical-harmonic space. Since $W(\vnhat)$ is typically
smooth, its spherical transform will be band-limited and straightforward
to handle.}. To
get around this problem, we choose to calculate the derivatives of
$W(\vnhat)$ in the flat-sky approximation where the differential
operators reduce to
\ba
\eth \approx -(\partial_x + i \partial_y), \\
\bar{\eth} \approx -(\partial_x - i \partial_y).
\ea
The derivatives are then trivially calculated on a regular Cartesian 
grid using finite differencing~\citep{smith07}. 

\begin{figure*}
  \centering
  \resizebox{0.55\textwidth}{!}{  
    \rotatebox{0}{\includegraphics{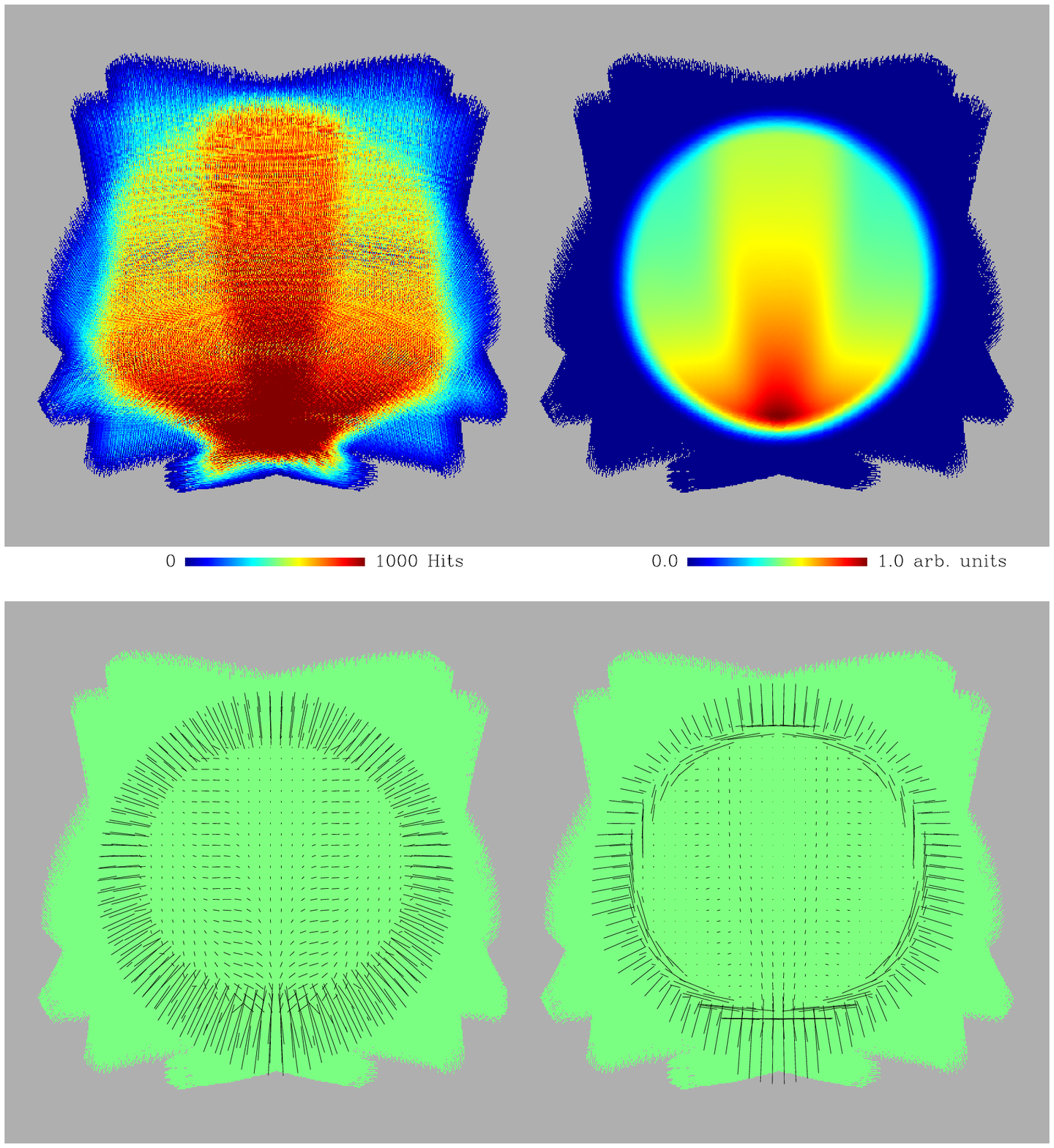}}}
  \caption{Inverse-variance weight functions used for power spectrum
  estimation for one of the southern fields. For the noise properties
  of our simulated data, the hit-map shown in the top left panel
  closely approximates the inverse-variance map. This map is heavily
  smoothed and apodized at the boundary of the map to produce the
  spin-0 weight function shown in the top right panel.
  The spin-1 and spin-2 weight functions, $\eth W$ and $\eth\eth W$,
  are shown (as vector fields) in the bottom left and right-hand
  panels respectively.}
  \label{fig:ppcl_weights}
\end{figure*}

The most optimal weighting scheme for a pseudo-$C_\ell$ analysis
involves different weight functions for each $C_\ell$ band-power
according to the signal-to-noise level expected in that band.
However, this is a costly solution (requiring $3 N_{\rm band}$
spherical harmonic transforms) and the indications are, from some
restricted tests that we have carried out, that the improvement in
resulting error-bars is small, at least for the specific noise
properties of our simulations. For the analysis presented here, we
have therefore chosen a simpler scheme whereby we have used a uniform
weight, appropriately apodized at the boundaries for the entire $\ell$
range for $E$-modes and for $\ell \le 200$ for $B$-modes. For $\ell >
200$ our simulated experiment is completely noise dominated for a
measurement of $B$-modes and so here we use an inverse-variance
weighting, again, appropriately apodized at the boundaries. For
simplicity, we have approximated the boundary of the map as a circle
of radius $11^{\circ}$. Note that restricting our power spectrum analysis
to this central region of our maps means we are effectively using only
$\sim 70$ per cent of the available data. To calculate the derivatives of the
weight functions, we project our weight maps (defined in \healpix)
onto a Cartesian grid using a gnomonic projection. Once the
derivatives of the weight maps have been constructed on the grid using
finite differencing, they are transformed back to the original
\healpix\, grid. An example of the inverse variance weight maps we
have used and the resulting spin-1 and spin-2 weight functions, $\eth
W$ and $\eth\eth W$, for one of our fields are shown in
Fig.~\ref{fig:ppcl_weights}.

\section{Results from simulations}
\label{sec:results}
The map-making and power spectrum estimation procedures described
above have been applied to each of our simulated datasets treating
each of our four observing fields independently. For any given
simulation set, we have 50 Monte-Carlo simulations so we can estimate the
uncertainties on the band-powers measured from each field. For each
realisation, we can then combine our measurements from the four fields
using inverse-variance weights to produce a final single estimate of
the power spectrum for each realisation. When presenting our results
below, in all cases, we plot the mean of these final estimates. For
our reference simulation, and for the $1/f$ noise systematics, the
error-bars plotted are calculated from the scatter among the
realisations and are those appropriate for a single realisation. When
investigating the $B$-mode bias from systematics, we plot the results
from signal-only simulations and the error-bars plotted are the
standard error on the mean. For some of the noise-related systematics,
we will also examine the impact of modulation in the map domain where
the effects are already clear.
\subsection{Reference simulation}
\begin{figure*}
  \centering
  \resizebox{0.55\textwidth}{!}{  
    \includegraphics{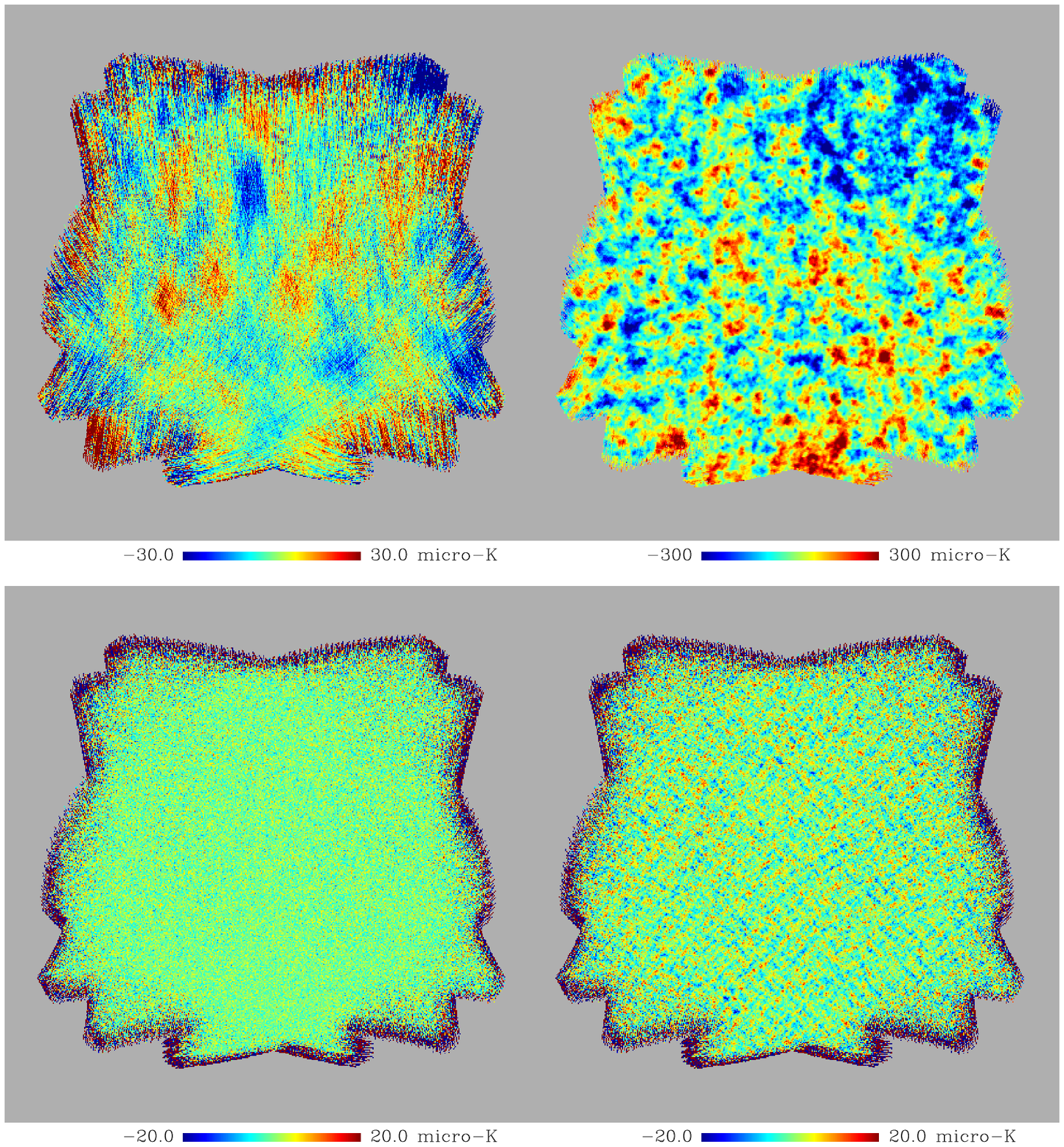}}
  \caption{Sample maps constructed from simulated time-stream
    containing noise only (left panels) and both signal and noise
    (right panels). Temperature maps are shown in the top panel and
    $U$-polarization maps are shown in the bottom panels. These maps
    are for one of our reference simulations with no explicit
    modulation scheme and no systematics included. Note the striping
    in the noise-only $T$ map which is completely absent from the $U$
    maps due to differencing of detector pairs before map-making.}
  \label{fig:maps_reference}
\end{figure*}
\begin{figure*}
  \centering
  \resizebox{0.70\textwidth}{!}{  
    \rotatebox{-90}{\includegraphics{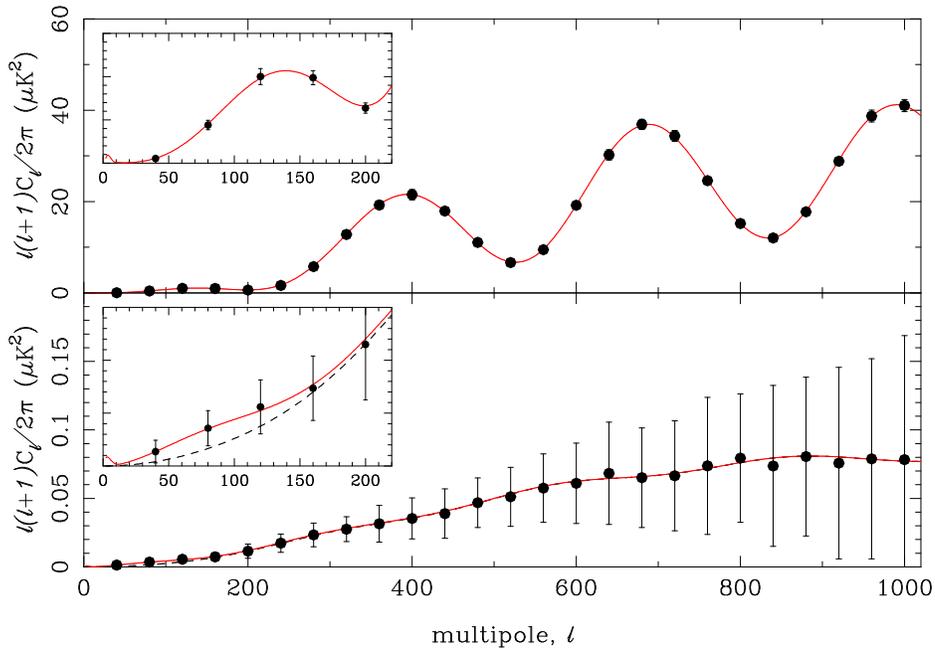}}}
  \caption{Mean recovered $E$-mode (top) and $B$-mode (bottom) power
  spectra for the reference simulations without explicit
  modulation. The errors plotted are those appropriate for a single
  realisation. The input CMB power spectra used to create the signal
  component of the simulations are shown as the red curves. In the
  bottom panel the total $B$-mode input signal (including lensing) for
  a tensor-to-scalar ratio of $r=0.026$ is shown as the red curve and
  the $B$-mode signal due to lensing alone is shown as the dashed
  curve. The $\ell < 200$ multipole range is shown in detail in the
  inset plots.}
  \label{fig:cls_reference}
\end{figure*}
To provide a reference for the results which follow, in
Figs.~\ref{fig:maps_reference} and \ref{fig:cls_reference} we show the results from our
suite of simulations with no explicit modulation and with no
systematics included. In Fig.~\ref{fig:maps_reference} we plot examples of the
reconstructed noise-only and signal-plus-noise $T$ and $U$ Stokes
parameter maps (the reconstructed $Q$ maps -- not shown -- are
qualitatively similar to $U$). The raw collecting power of an
experiment like \clover\, is apparent from the top two panels in this
figure. Although the noise $T$ map shown in the top-left panel is
clearly dominated by striping due to the correlated noise from the
atmosphere, only signal is apparent in the signal-plus-noise map shown in
the top-right panel. (In fact, the noise contribution to the $T$
signal-plus-noise map is significant, particularly on large scales, and
so would need to be accounted for when measuring the temperature power
spectrum.) Conversely, the $U$ noise map is dominated by
white noise; the correlated component of the atmosphere has been
removed completely from the polarization time-streams (as has the $T$
sky signal) by differencing detector pairs before map-making.

Figure~\ref{fig:cls_reference} shows the mean recovered $E$- and $B$-mode power
spectra from our reference simulations for the case of no explicit
modulation. Here, we see that our analysis is unbiased and recovers the input
polarization power spectra correctly. For an input tensor-to-scalar
ratio of $r=0.026$, we recover a detection of $B$-modes \emph{in
excess} of the lensing signal of $1.54\sigma$.
(We argue in Section~\ref{sec:fisher} that this is an under-estimate of
the detection significance by around 10 per cent due to our ignoring small
anti-correlations between the errors in adjacent band-powers.)

The corresponding plots for the stepped and continuously rotating HWP
are very similar apart from the reconstructed polarization maps at the
very edges of the fields where a modulation scheme increases the
ability to decompose into the $Q$ and $U$ Stokes parameters. Since the
edges of the field are excluded in our power spectrum analysis in any
case (see Fig.~\ref{fig:ppcl_weights}), we find that the performance
(in terms of $C_\ell$ errors) of all three types of experiments which
we have considered is qualitatively the same in the absence of
systematic effects. Note that for all the systematics we have
considered, the effects on the recovery of the $E$-mode spectrum is
negligible and so, in the following sections, we plot only the
recovered $B$-mode power spectra which are the main focus of this
paper.

\subsection{$1/f$ detector noise}

\begin{figure*}
  \centering
  \resizebox{0.9\textwidth}{!}{  
    \rotatebox{-90}{\includegraphics{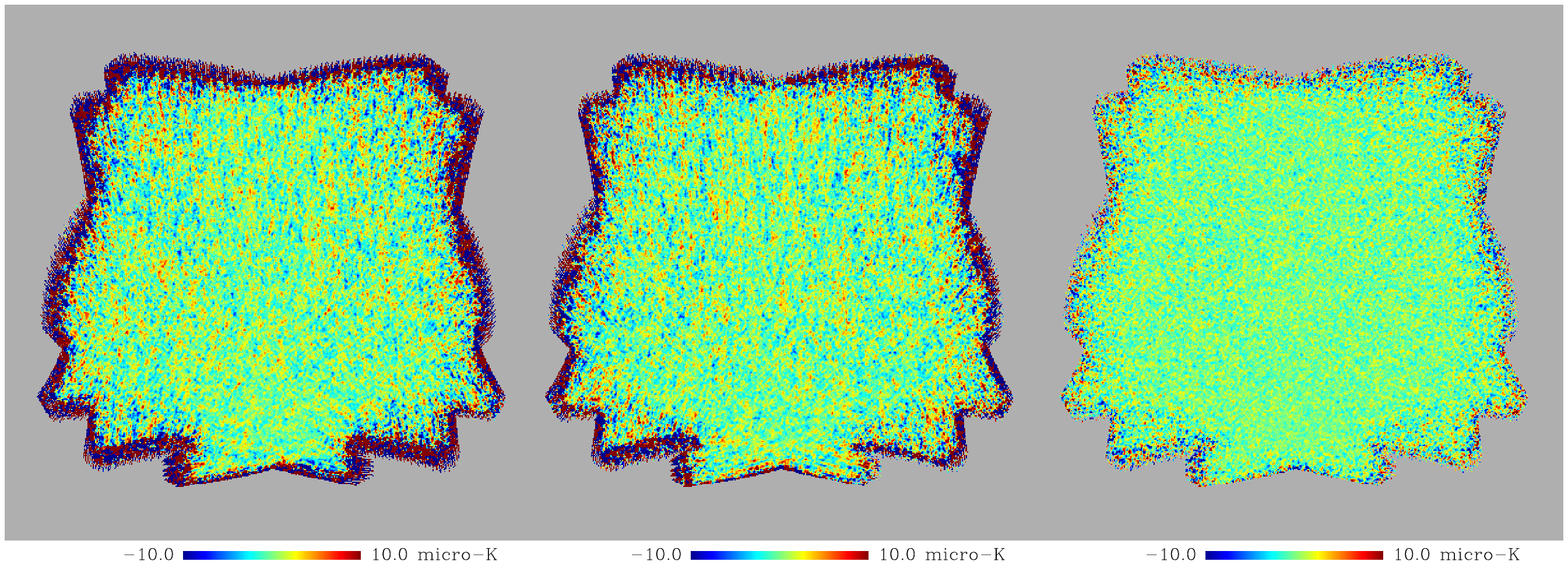}}}
  \caption{Sample noise-only $U$-maps from simulations containing a
  $1/f$ component correlated to the detector noise}. For display
  purposes only, the maps have been smoothed with a Gaussian with a FWHM
  of 7 arcmin. In the case where explicit modulation is either absent (left
  panel) or slow (stepped HWP; middle panel), the $1/f$ noise
  results in faint residual stripes in the polarization maps. In the case of
  a continuously rotating HWP, the polarization signal is modulated
  away from the low frequency $1/f$ resulting in white-noise behaviour
  in the polarization map (right panel).
  \label{fig:maps_det_noise}
\end{figure*}

Figure~\ref{fig:maps_det_noise} shows the recovered noise-only maps from
a simulation containing a correlated $1/f$ detector noise
component with a knee frequency of $f_{\rm knee} = 0.1$~Hz. 
In this figure, we have plotted the noise-only maps from
the three types of experiment we have considered: no modulation; a
HWP which is stepped by 20$^{\circ}$ at the end of each azimuth scan; and
a HWP continuously rotating at $3$~Hz. The impact of modulation on $1/f$
detector noise is clear from this plot -- as described in
Section~\ref{sec:modulation}, the continuously rotating HWP shifts
the polarization band in frequency away from the $1/f$ detector noise
leaving only white noise in the resulting map. A stepped HWP, on the
other hand, does not mitigate $1/f$ detector noise in this way
and so noise striping is apparent in the middle panel of
Fig.~\ref{fig:maps_det_noise}. 

\begin{figure*}
  \centering
  \resizebox{0.9\textwidth}{!}{  
    \rotatebox{-90}{\includegraphics{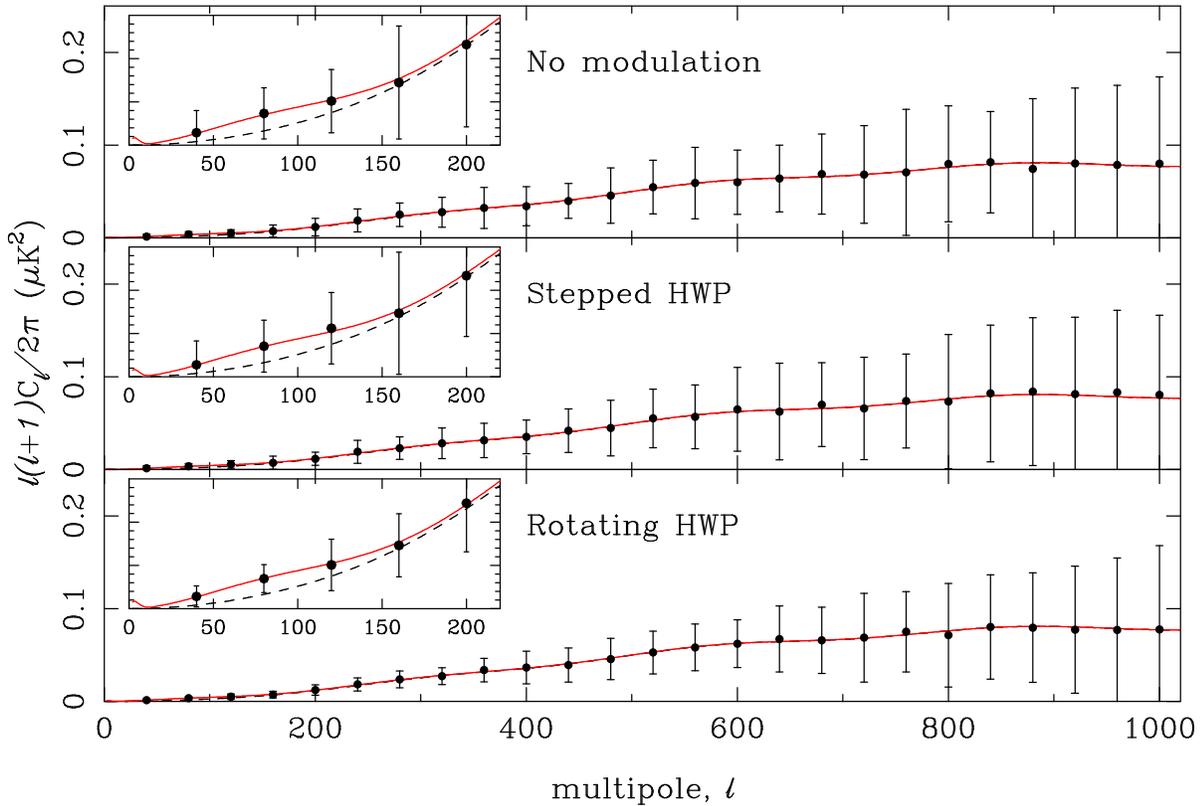}}}
  \caption{Recovered $B$-mode power spectra for the simulations
  including a correlated $1/f$ component to the detector noise with
  $f_{\rm knee} = 0.1$~Hz for no modulation (top), a stepped HWP
  (middle) and a HWP continuously rotating at 3~Hz (bottom). See Table
  \ref{tab:simsummary} for the significances with which each
  experiment detects the input signals.}
  \label{fig:cls_det_noise}
\end{figure*}

The $B$-mode power spectra measured from our signal-plus-noise simulations
including $1/f$ detector noise are shown in
Fig.~\ref{fig:cls_det_noise} (again for $f_{\rm knee} = 0.1$~Hz) where
we show the results from all three types of experiment. 
Examination of the figure suggests that the presence of a $1/f$
component in the detector noise leads to a significant degradation in
the ability of the unmodulated and stepped-HWP experiments to recover
the input $B$-mode signal. This degradation happens at all multipoles
but is particularly acute on the largest scales ($\ell < 200$) where
the primordial $B$-mode signal resides. The marginal detection of
primordial $B$-modes (for $r = 0.026$) which we saw in our reference
simulation (Fig.~\ref{fig:cls_reference}) is now completely destroyed
by the presence of the $1/f$ correlated detector noise. Furthermore,
note that our analysis of the simulated data sets is optimistic in the
sense that we have assumed that any correlated noise can be modelled
accurately -- that is, our noise-only simulations, which we use to
measure the noise bias, are generated from the same model noise power
spectrum used to generate the noise component in our signal-plus-noise
simulations. (This is the reason that our recovered spectra are
unbiased.) However, for the analysis of a real experiment, the noise
properties need to be measured from the real data and there are
uncertainties and approximations inherent in this process. Any
correlated detector noise encountered in a real experiment is unlikely
to be understood to the level which we have assumed in our analysis
and so will likely result not only in the increased uncertainties we
have demonstrated here but also in a biased result at low
multipoles. Estimating cross-spectra between maps made from subsets of
detectors for which the $1/f$ detector noise is measured to be
uncorrelated is a simple way to avoid this noise bias issue, at the
expense of a small increase in the error-bars~\citep{hinshaw03}.

The results from the simulations where we have continuously modulated
the polarization signal recover the input $B$-mode signal to the same
precision that we saw with our reference simulation -- our marginal
detection of $r=0.026$ is retained even in the presence of the
correlated detector noise. In Section \ref{sec:significances} and
Table \ref{tab:simsummary} we show quantitatively that, for a detector
knee frequency of $0.1$~Hz, the significance with which the
continuously modulated experiment detects the primordial $B$-mode
signal is roughly twice that found for the un-modulated and
stepped-HWP experiments. Also detailed in Table \ref{tab:simsummary}
are the results from our $1/f$ noise simulations with knee frequencies
of $0.05$ and $0.01$~Hz. We see, as expected, that the impact of fast
modulation is less for a lower knee frequency --- for $f_{\rm knee} =
0.05$, rapid modulation still significantly out-performs the
un-modulated and stepped-HWP experiments while for $f_{\rm knee} =
0.01$~Hz, there is essentially no difference between the performance
of the three types of experiment.

Note that, in the case of rapid modulation, because the polarization
signal is moved completely away from the $1/f$ frequency regime, the
recovered spectra should be immune to the issues of mis-estimation or
poor knowledge of the noise power spectrum at low frequencies
mentioned above. Although detector $1/f$ noise can be mitigated by
other methods (e.g. using a more sophisticated map-maker;
\citealt{sutton09}), these usually require accurate knowledge of the
low-frequency noise spectrum unlike the hardware approach of fast
modulation.

\subsection{Polarized atmospheric $1/f$}
In contrast to the addition of $1/f$ detector noise, which can be
successfully dealt with by rapid modulation, all three types of
experiment are degraded similarly by polarized low-frequency noise in
the atmosphere. In particular, the errors at low multipoles are
inflated by a large factor since the large amount of polarized
atmosphere which we have input to the simulations swamps the input
$B$-mode signal for $r=0.026$. We stress that the levels of polarized
atmosphere we have used in these simulations were deliberately chosen to
demonstrate the point that modulation does not help and the levels are
certainly pessimistic. 

\subsection{Calibration errors}
\begin{figure*}
  \centering
  \resizebox{0.90\textwidth}{!}{  
    \rotatebox{-90}{\includegraphics{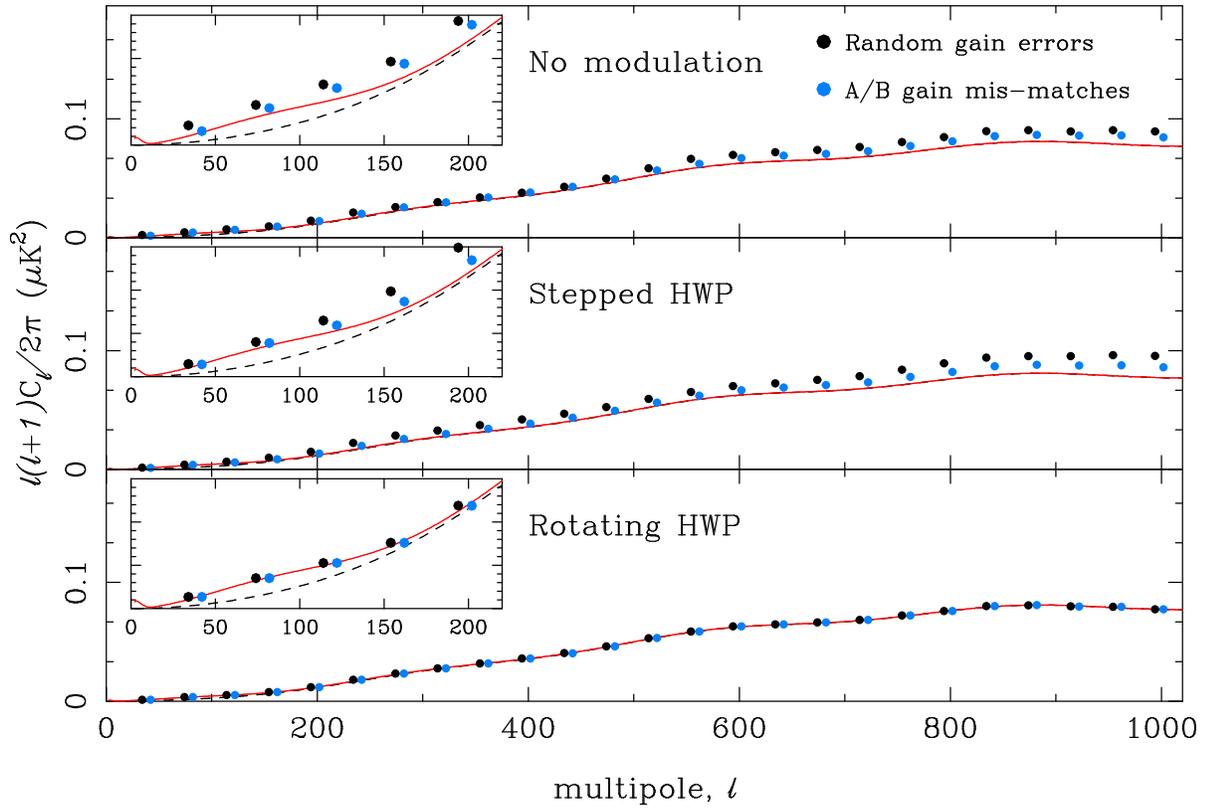}}}
  \caption{Mean recovered $B$-mode power spectra for the simulations
  including random gain errors across the focal plane (black points) or
  systematic 1 per cent A/B gain mis-matches between detector pairs (blue points)
  for no modulation (top), a stepped
  HWP (middle) and a continuously rotating HWP (bottom). These spectra
  are measured from our suite of signal-only simulations. Our
  simulations containing both signal and noise exhibit the same biased
  recovery for the no-modulation and stepped-HWP cases. The presence
  of a fast modulation scheme (bottom panel) mitigates entirely the
  bias caused by these gain mis-matches. The standard errors in these
  mean recovered spectra are smaller than the plotted symbols.}
  \label{fig:cls_gain_errors}
\end{figure*}

The power spectra recovered from signal-only simulations where we
introduced random gain errors (constant in time) across the focal plane,
or 1 per cent systematic A/B gain mis-matches between the two detectors
within each pixel are shown in Fig.~\ref{fig:cls_gain_errors}. In both
cases, we see a clear bias in the recovered $B$-mode signal in the
absence of fast modulation, but the bias is mitigated entirely by the
presence of a HWP rotating at 3~Hz. The bias is also mitigated to some
degree by the stepped HWP but not completely. In our simulations, the bias is
generally larger for the case of random gain errors since the
variance (across the focal plane) of the gain mismatches is twice as large
in the former case. For our simulations where we allowed detector gains to drift
over the course of a two-hour observation, we found a similarly
behaved $B$-mode bias to those shown in Fig.~\ref{fig:cls_gain_errors}
but with a smaller magnitude (due to the two-hour drifts averaging down
over the eight-hour observation).

A mis-match between the gains of the two detectors within a pixel
corresponds to a $T \rightarrow Q$ leakage in the detector
basis. The projection of this instrumental polarization onto the sky
will therefore be suppressed if a wide range of sensitivity directions
$\phi_i$ contribute to each sky pixel, as is the case for fast
modulation. Note that in the case of a stepped HWP, one should be
careful to design the stepping strategy in such a way that it does not
undo some of the effect of sky rotation. During our analysis, we have
found that the performance of a stepped HWP in mitigating systematic
effects can depend critically on the direction, magnitude and
frequency of the HWP step applied. In fact, for some set-ups we have
investigated, a stepped HWP actually worsened the performance in
comparison to the no modulation case due to interactions between the
scan strategy and HWP stepping strategy. However, the results plotted in
Fig.~\ref{fig:cls_gain_errors} for the stepped HWP case are for a
HWP step of $20^{\circ}$ between each azimuth scan which is large and
frequent enough to ensure that such interactions between the stepping strategy
and the scan strategy are sub-dominant. 

\subsection{Mis-estimated polarization angles}
The next set of systematics we have considered concern a
mis-estimation of both the detector orientation angles (i.e.\ the
direction of linear polarization to which each detector is sensitive
to) and, for the case where a HWP is employed, a mis-estimation of the
HWP orientation. We have performed simulations including both a random
scatter (with an RMS of $0.5^{\circ}$) and a systematic offset of
$0.5^{\circ}$ in the simulated detector and HWP angles. Note that for
the systematic offset in the detector angles, the same offset is
applied to all detectors. For both the detector angles and the HWP,
the offset introduced corresponds to a systematic error in the
estimation of the global polarization coordinate frame of the
experiment and the effects are therefore degenerate.

For the simulations which included a random scatter in the angles
(both detector angles and HWP orientation), we found neither a bias in
the recovered $B$-mode power spectra, nor a degradation in the
error-bars from the simulations containing both signal and
noise. Following the discussion in Section~\ref{sec:systematics},
common errors in the detector angles for the pair of detectors in a single
focal-plane pixel give rise to a rotation of the polarization sensitivity
direction of the pixel, while differential errors reduce the polarization
efficiency.
For a typical differential scatter of $\sqrt{2}\times 0.5^{\circ}$,
the reduction in the polarization efficiency ($\sim 10^{-4}$) is
negligible. For a given pixel on the sky, the impact of the polarization
rotation is suppressed by $\sqrt{N_\mathrm{sample}}$, where
$N_\mathrm{sample}$ is the total number of samples contributing to that
pixel with independent errors in the angles. The combination of a
large number of detectors and, in the case of random HWP angle errors,
their assumed short correlation time in our simulations renders the
effect of small and random scatter in the angles negligible.

\begin{figure}
  \centering
  \resizebox{0.48\textwidth}{!}{  
    \rotatebox{-90}{\includegraphics{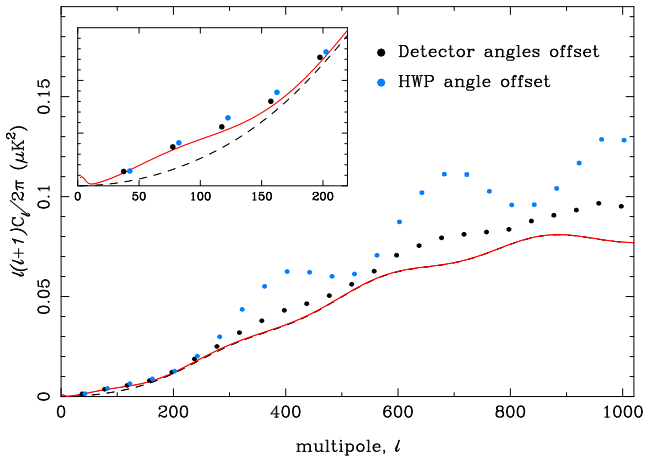}}}
  \caption{Mean recovered $B$-mode power spectra for the signal-only simulations
  including mis-estimated detector polarization sensitivity angles (black
   points) and mis-estimated HWP angles (blue points) where the angles have been
  systematically offset by 0.5$^{\circ}$ in both cases. The standard errors in these
  mean recovered spectra are smaller than the plotted symbols.} 
  \label{fig:cls_pol_angles}
\end{figure}

The results from simulations which included a systematic error in the
angles are shown in Fig.~\ref{fig:cls_pol_angles}, where we plot the recovered
$B$-mode power spectrum from our signal-only simulations. In contrast
to the simulations with random scatter, there is a clear mixing between $E$
and $B$ due to the systematic mis-calibration of the polarization
coordinate reference system of the instrument. A global mis-estimation
of the polarization direction by an angle $\psi$ in the reconstructed maps
leads to spurious $B$-modes with
\be
C_\ell^B = \sin^2(2\psi) C_\ell^E \approx 4 \psi^2 C_\ell^E .
\ee
Note that in
Fig.~\ref{fig:cls_pol_angles}, we show the results from the detector angle
systematic only for the case of the non-modulated experiment but the
plot is identical for both the stepped and continuously rotating HWP --
polarization modulation cannot mitigate a mis-calibration of detector 
angles. The fact that the mixing apparent in Fig.~\ref{fig:cls_pol_angles} is
greater for the HWP mis-calibration is simply because rotating the
waveplate by $\psi$ rotates the polarization direction by $2\psi$.
Although the spurious $B$-mode power is most
noticeable at high multipoles, where $\ell(\ell+1)C_\ell^E$ is largest,
it is also present on large scales and,
as is clear from the plot, would bias a measurement of the $B$-mode 
spectrum at all multipoles. 

\subsection{Mis-estimated time-constants}

The power spectra measured from simulations which included random and
systematic errors in the detector time-constants displayed neither a
bias, nor a degradation in error-bars. This was to be expected for the
slow scan speed and extremely fast time-constants we have considered
in this analysis -- the response function of the \clover\, detectors
is effectively phase-preserving with zero attenuation in the frequency
band which contains the sky-signal in our simulations. Note that this
would not necessarily have been the case had we considered a much
faster scan speed or more rapid polarization modulation.

\subsection{Pointing errors}

Our analysis of simulations where we have introduced a jitter in the
pointing and/or an overall wander in the pointing suggest that these
systematics have only a very small effect on the recovered $B$-mode
power spectra, at least for the levels which we have considered
(i.e. a $30$ arcsec random jitter in the pointing and/or an overall
wander of the pointing by $1$ arcmin over the course of a two-hour
observation). The only observed effect was a slight suppression of the
recovered $B$-mode signal at high multipoles consistent with a
slight smearing of the effective beam. We note however that the effect 
we observed was extremely small and was only noticeable in our
signal-only simulations. For our simulations containing noise, the
effect was completely swamped by the errors due to random noise. 

In principle, pointing errors can also lead to leakage
from $E$ to $B$~\citep{hu03,odea07}. In Appendix~\ref{app:pointing}
we develop a toy-model for
the leakage expected from random pointing jitter in the case of
a scan/modulation strategy that produces a uniform spread of polarization
sensitivity directions in each sky pixel (such as by fast modulating
with a HWP). The result is a white-noise spectrum of $B$-modes but,
for the simulation parameters adopted here, the effect is very small --
less than $1$ per cent of the $B$-mode power induced by weak gravitational lensing
on large scales.

\subsection{Differential transmittance in the HWP}
The power spectra reconstructed from our simulations which 
included a 2 per cent differential transmittance in the HWP exhibited
no degradation in the accuracy of the recovered $B$-mode
signal --- even the relatively simple recipe which we have used to
remove the HWP-synchronous signals from the time-stream 
(see Section \ref{sec:map-making}) appears sufficient to recover the
$B$-mode signal to the same accuracy as was seen in our reference
simulations. (We quantify this statement in the next section where we
estimate the detection significances with which the different simulations
detect the $E$ and $B$-mode signals). As mentioned in
Section~\ref{sec:modulation}, the recovered polarization signal is
mis-calibrated by $\sim 2$ per cent in amplitude ($4$ per cent in
power). Compared to the random noise however, this mis-calibration is
a small effect and is easily dealt with during, e.g. a cosmological
parameter analysis by marginalising over it. 

Note that no prior information on the level of differential
transmittance was used during our analysis of the data. Our technique
for removing the HWP-synchronous signals is a blind one in this
sense and should work equally well for other HWP-systematic effects
that result in spurious signals at harmonics of the HWP rotation
frequency, $f_\lambda$.

\section{Discussion}
\label{sec:discussion}

\subsection{Controlling systematics with polarization modulation}
\label{sec:significances}
The main goal of the analysis presented in this paper is to
demonstrate and quantify with simulations the impact of two types of
polarization modulation (slow modulation using a stepped HWP and rapid
modulation with a continuously rotating HWP) on the science return of
upcoming CMB $B$-mode experiments in the presence of various
systematic effects. Although our list of included systematics is not
an exhaustive one (in particular, we are still investigating the case
of imperfect optics), we are nevertheless in a position to draw some
rather general conclusions regarding the usefulness of modulation in
mitigating systematics. It is, of course, important to bear in mind
that we have only considered two examples of a HWP-related systematic
effect (imperfect HWP angles and differential transmittance in the
plate). The are many more possible effects which will need to be
well understood and strictly controlled if fast polarization
modulation with HWPs is to realise its potential.

\subsubsection*{(i) Systematics mitigated by modulation}
\begin{itemize} 
\item{\bf Correlated $1/f$ detector noise:} As expected by the general
  reasoning of Section~\ref{sec:modulation}, and further borne out
  by our results from simulations, rapid polarization modulation is
  extremely powerful at mitigating a correlated $1/f$ component in the
  detector noise. Any such $1/f$ component is not mitigated by
  a HWP operating in stepped mode. 
\item{\bf Calibration errors:} Our results demonstrate that fast
  modulation is also useful for mitigating against possible
  calibration errors since it greatly increases the range of
  directions over which sky polarization is measured in a given pixel.
  For example, the clear bias introduced in our simulations by random
  gain drifts or systematic mis-calibrations between detectors was
  mitigated entirely by the HWP continuously rotating at $3$~Hz. This
  bias was also partly (but not completely) mitigated by stepping the
  HWP by 20$^{\circ}$ between each of our azimuth scans. For some stepping
  strategies we have investigated, the bias actually increased --- a
  poor choice of stepping strategy can actually be worse than having
  no modulation because of interactions between the sky rotation and
  the HWP orientations.
\end{itemize}

\begin{table*}
\caption{Detection significances (in units of $\sigma$) for our
  reference simulations, for our simulations with $1/f$ noise
  systematics and for our simulations with a 2 per cent differential
  transmittance in the HWP. Also included for comparison are the predicted
  detection significances from a Fisher matrix analysis of the power
  spectrum errors (see Section \ref{sec:fisher}) and from the
  simulations containing isotropic and uniform Gaussian noise (see
  text). The rightmost column displays the significance of the
  detection of the $B$-mode signal in excess of the lensing signal
  which corresponds directly to the significance with which each
  simulation detects the input tensor-to-scalar ratio of $r=0.026$.}
\begin{center}
\begin{tabular}{c|c|c|c|c}
Simulation                   & Modulation   & $E$-mode & $B$-mode & $r = 0.026$ \\
\hline
Fisher predictions           & ---          & $128.6$ & $10.24$ & $1.90$ \\
\hline
Uniform noise                & ---          & $126.4$ & $10.30$ & $1.45$ \\
\hline
Reference simulation         & None         & $127.8$ & $10.41$ & $1.54$ \\
                             & Stepped HWP  & $130.1$ & $10.11$ & $1.41$ \\
                             & Rotating HWP & $127.7$ & $10.72$ & $1.45$ \\
\hline
$1/f$ detector noise         & None         & $124.2$ & $7.29$  & $0.83$ \\
($f_{\rm knee} = 0.1$~Hz)       & Stepped HWP  & $124.6$ & $7.09$  & $0.79$ \\
                             & Rotating HWP & $126.8$ & $9.96$  & $1.47$ \\
\hline
$1/f$ detector noise         & None         & $125.1$ & $8.61$  & $0.95$ \\
($f_{\rm knee} = 0.05$~Hz)      & Stepped HWP  & $127.5$ & $8.59$  & $1.14$ \\
                             & Rotating HWP & $125.6$ & $10.31$ & $1.45$ \\
\hline
$1/f$ detector noise         & None         & $126.8$ & $9.78$  & $1.30$ \\
($f_{\rm knee} = 0.01$~Hz)      & Stepped HWP  & $128.0$ & $9.99$  & $1.44$ \\
                             & Rotating HWP & $127.1$ & $10.28$ & $1.40$ \\
\hline
Polarized atmosphere         & None         & $122.9$ & $6.86$  & $0.12$ \\
                             & Stepped HWP  & $124.9$ & $6.99$  & $0.10$ \\
                             & Rotating HWP & $126.0$ & $8.24$  & $0.21$ \\
\hline
Differential transmittance   & Rotating HWP & $127.6$ & $10.92$ & $1.59$ \\
\hline
\end{tabular}
\end{center}
\label{tab:simsummary}
\end{table*}

\vspace{-6mm}
\subsubsection*{(ii) Systematics not mitigated by modulation}
\begin{itemize}
\item{\bf Polarized $1/f$ atmosphere:} No amount of modulation (rapid or
  slow) will mitigate a polarized $1/f$ component in the
  atmosphere. The results from our simulations containing polarized
  atmosphere are summarised in Table \ref{tab:simsummary}. 
\item{\bf Pointing errors:} For our simulations which included pointing
  errors, the effect on the recovered $B$-mode power spectra was
  extremely small and was equivalent to a slight smoothing of the
  effective beam. Although the same amount of smoothing was observed
  in all the simulations (and so the effect is not mitigated by
  polarization modulation), the effect is negligible for the
  sensitivities and beam sizes considered here. The further leakage of
  $E$-mode power into $B$ modes due to pointing errors was, as expected
  (see Appendix~\ref{app:pointing}) unobservably small in our simulations.
\item{\bf Mis-calibration of polarization angles: } A polarization
  modulation scheme does not mitigate a systematic error in the
  calibration of the polarization sensitivity directions. Experiments
  using a HWP will require precise and accurate measurements of the
  HWP angle at any given time to avoid the $E \rightarrow B$ mixing
  apparent in Fig.~\ref{fig:cls_pol_angles}.
\end{itemize}

Our results are in broad agreement with those of a similar study by
\cite{mactavish08} who based their analysis on signal-only simulations
of the \spider\, experiment. Both \cite{mactavish08} and this study find
that polarization modulation with a continuously rotating HWP is
extremely effective in mitigating the effects of $1/f$ detector noise
but that, in the presence of significant $1/f$ noise, the
analysis of an experiment where modulation is either absent or slow 
will require near-optimal map-making techniques. In addition, both studies find that the effect
of small and random pointing errors on the science return of upcoming
$B$-mode experiments is negligible given the experimental
sensitivities. The two analyses also find that the effect of random errors
(with $\sim 0.5^{\circ}$ RMS) in the detector polarization sensitivity
angles is negligible but that the global polarization coordinate frame
of the experiment needs to be measured carefully --- \cite{mactavish08}
quote a required accuracy of $< 0.25^{\circ}$ for \spider\, which is
consistent with the requirement for an unbiased measurement of
the $B$-mode signal (for $r = 0.026$) at $\ell < 300$ with \clover. Finally, both
studies suggest that, in the absence of fast modulation, relative gain errors
will also need to be controlled to the $< 1$ per cent level (although, in
this paper, we have demonstrated that such gain errors are almost
entirely mitigated using a fast modulation scheme; see
Fig.~\ref{fig:cls_gain_errors}).  

In Table~\ref{tab:simsummary}, we quantify the impact of modulation on
the $1/f$ noise systematics we have considered in this work by
considering the significances with which we detect the $E$-mode and
$B$-mode signals. For comparison, the detection significances in the
presence of a 2 per cent differential transmittance in the HWP are also
presented. To calculate the total significance of the detection
we compute the Fisher error on the amplitude of a fiducial
spectrum,
\be
\frac{S}{N} = \left( \sum_{bb'} P_b^{\rm fid} {\rm cov}^{-1}_{bb'}
P_{b'}^{\rm fid} \right)^{1/2},
\ee
where ${\rm cov}^{-1}_{bb'}$ is the inverse of the band-power covariance
matrix for the given spectrum.
For the total significance of a detection of $E$ or $B$-modes, the
fiducial band-powers are simply the binned input power spectra. In
order to estimate the significance of a detection of primordial
$B$-modes, we subtract the lensing contribution from the input
$B$-mode power spectra. Because the primordial $B$-mode power spectrum
is directly proportional to the tensor-to-scalar ratio, $r$, the
significance with which we detect the $B$-mode signal in excess of the
lensing signal translates directly to a significance for the detection
of our input tensor-to-scalar ratio of $r=0.026$.
When analysing the results of the
simulations, we approximate ${\rm cov}^{-1}_{bb'} \approx \delta_{bb'}/
\sigma_b^2$ since we are unable to estimate the off-diagonal elements
from our small number of realisations (50) in each simulation set.
We know from a Fisher-based analysis (see Section~\ref{sec:fisher}),
the results of which are also reported in Table~\ref{tab:simsummary},
that neighbouring band-powers on the largest scales are, in fact,
$\sim 10$ per cent 
anti-correlated, and the diagonal approximation therefore
\emph{underestimates} the detection significance by $\sim 10$ per cent.
For consistency, the numbers quoted for the Fisher analysis ignore the
off-diagonal elements of the covariance matrix. Including the correlations
increases the $E$-mode significance to $144.6$  (from $128.6$),
the total $B$-mode significance to $11.57$ (from $10.24$) and
the primordial $B$-mode significance to $2.04$ (from $1.90$).

In comparing the entries in Table~\ref{tab:simsummary} one should
keep in mind that the significances reported for the simulations are
subject to a Monte-Carlo error due to the finite number ($N_{\mathrm{sim}}=50$)
of simulations used to estimate the band-power errors. Approximating the
band-powers as uncorrelated and Gaussian distributed, the
sampling error in our estimates of the $S/N$ is
\be
\Delta (S/N) \approx \frac{1}{S/N} \frac{1}{\sqrt{2N_{\mathrm{sim}}}}
\left[\sum_b \left(\frac{P_b^{\mathrm{fid}}}{\sigma_b}\right)^4\right]^{1/2} .
\ee
For the reference simulation, this gives an error of $0.15$ in the
significance of a detection of $r$ and $0.22$ in the significance of the
total $B$-mode spectrum. The size of these errors likely explain the
apparent anomalies that rotating the HWP degrades the detection of $r$
in the reference simulation, and that adding $1/f$ detector noise
improves the detection of $r$ over the reference simulation for the case
of a rotating HWP.

Also included in Table~\ref{tab:simsummary} is the performance of our
experiment as estimated from a set
of simple map-based simulations where we have injected uniform and
isotropic white noise into signal-only $T$, $Q$ and $U$ maps
directly. In these simulations, and also for the Fisher analysis,
the white-noise levels were chosen to
match the noise levels in our main analysis and so they have identical
raw sensitivity to the time-stream simulations but with perfectly
behaved noise properties. The broad agreement between our full
time-stream simulations and these simple map-based simulations
suggests that the anisotropic noise distribution introduced by the
\clover\, scan strategy does not have a large impact on the
performance of the experiment. This agreement also suggests that the
slightly poorer performance of the simulations in recovering the
$r=0.026$ primordial $B$-mode signal as compared to the Fisher
predictions is due to the sub-optimal performance on large scales of the
(pure) pseudo-$C_\ell$ estimator we have used.

\subsection{Importance of combining data from multiple detectors}
\label{sec:differencing}
For all of our analyses up to this point, in order to remove the
correlated $1/f$ atmospheric noise from the polarization
analysis, we have differenced detector pairs before
map-making. However, as mentioned in Section~\ref{sec:modulation}, for
the case of a continuously modulated experiment, it is possible to
measure all three Stokes parameters from a single detector in
isolation. Here, we argue that this may be a poor choice of analysis
technique in the presence of a highly correlated and common-mode
systematic such as atmospheric $1/f$, at least when one employs
real-space demodulation techniques such as those that we have used in
this analysis. The key point to appreciate here is that, even with a
rapid modulation scheme, and for an ideal experiment, it is impossible
to separate completely the temperature and polarization signals in
real space using only a single detector.\footnote{We note that this is
not necessarily true for the case of genuinely band-limited
temperature and polarization signals when a classical lock-in
technique (such as that used in the analysis of the \maxipol\, data;
\citealt{johnson07}) is used to perform the demodulation. We are
currently working to integrate such a technique into our analysis.}
In contrast, the technique of detector differencing achieves this
complete separation of the temperature and polarization signals (again
for the case of an ideal experiment). Note that this is true even in
the signal-only case. Consider again the modulated signal, in the
absence of noise, from a single detector sensitive to a single
polarization: \be d_i = \left[ T(\theta) + Q(\theta) \cos(2\phi_i) +
U(\theta)\sin(2\phi_i) \right]/2.  \ee If for each observed data
point, $d_i$, the true sky signals, $T$, $Q$ and $U$ are different (as
is the case for a scanning experiment), there is clearly no way to
recover the true values of $T$, $Q$ and $U$ at each point in time.
The approximation that one must make in order to demodulate the data
in real space goes to the very heart of map-making -- that the true
continuously varying sky signal can be approximated as a pixelised
distribution where $T$, $Q$ and $U$ are taken to be constant within
each map-pixel. Armed with this assumption, all three Stokes
parameters can be reconstructed from a single detector time-stream
using a generalisation of equation~(\ref{eqn:qu_mapmaking}):
\be
\left( \begin{array}{c} T \\ Q \\ U \end{array} \right) = 2 \, {\mathsf M}^{-1} \cdot
\left( \begin{array}{c}
    \lgl d_i \rgl \\
    \lgl\cos(2\phi_i)d_i\rgl \\
    \lgl\sin(2\phi_i)d_i\rgl \\ \end{array} \right).
\label{eqn:tqu_mapmaking}
\ee
where the decorrelation matrix, ${\mathsf M}$ is now given by
\be 
{\mathsf M} = \left( \begin{array}{ccc} 
    \!\!\! 1 &  \!\!\! \lgl\cos(2\phi_i)\rgl &  \!\!\!\!\! \lgl\sin(2\phi_i)\rgl \\
    \!\!\! \lgl\cos(2\phi_i)\rgl &  \!\!\! \lgl\cos^2(2\phi_i)\rgl &
    \!\!\!\!\! \lgl\cos(2\phi_i)\sin(2\phi_i)\rgl \\
    \!\!\! \lgl\sin(2\phi_i)\rgl &  \!\!\! \lgl\cos(2\phi_i)\sin(2\phi_i)\rgl &
    \!\!\!\!\! \lgl\sin^2(2\phi_i)\rgl \\ \end{array} \!\!\!\! \right).
\ee
\begin{figure}
  \centering
  \resizebox{0.48\textwidth}{!}{  
    \rotatebox{-90}{\includegraphics{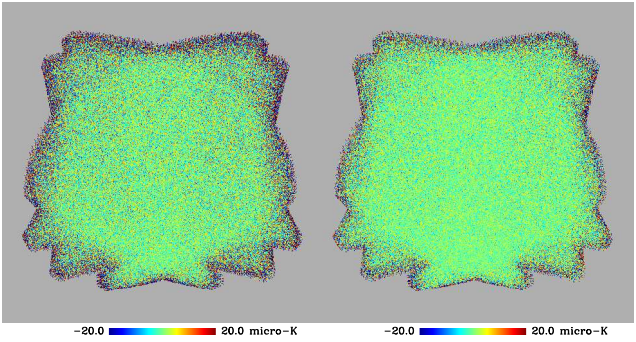}}}
  \caption{Recovered noise only $U$-polarization maps from one of our
  reference simulations with continuous modulation. The map on the
  left is reconstructed from demodulated single detector ``pure'' $U$
  time streams and has not used information from multiple detectors to
  separate the $T$ and polarization signals. The map on the right is
  made from demodulated detector-pair ``pure'' $U$ time-streams and
  explicitly combines information from the two detectors within each
  pixel to separate $T$ from $Q/U$. Although striping is absent from
  both maps, the white-noise level in the detector-differenced map is
  reduced compared to that made using the non-differencing
  analysis.}
  \label{fig:maps_diff_nodiff}
\end{figure}
If, on the other hand, the data from different detectors are combined
(e.g.\ when detector differencing is used), the situation is different
-- because the two detectors within a pixel observe exactly the same
un-polarized component of the sky signal at exactly the same time,
differencing the detectors removes
the $T$ signal completely without any assumptions regarding the scale
over which the true sky signal is constant. In this case, the
decorrelation of $Q$ and $U$ using equation~(\ref{eqn:qu_mapmaking})
still requires an assumption regarding the constancy of the $Q$ and
$U$ signals over the scale of a map-pixel but now the much larger
temperature signal has been removed from the polarization analysis
completely. 

Now, if in addition to the sky signal, we have a common-mode
time-varying systematic such as an un-polarized $1/f$ component in the
atmosphere, this contaminant will again be removed entirely with the
detector differencing technique (as long as it is completely
correlated between the two detectors) whilst it will introduce a
further approximation into any attempt to decorrelate all three Stokes
parameters from a single detector time-stream using
equation~(\ref{eqn:tqu_mapmaking}). 

To illustrate this point, and to stress the importance of combining
data from multiple detectors, we have re-analysed the simulated data
from our set of continuously modulated reference simulations but now
we perform the demodulation at the time-stream level for either single
detectors or single detector-pairs in isolation. Firstly, we
demodulate each detector time-stream individually using
equation~(\ref{eqn:tqu_mapmaking}) but now the averaging is performed
over short segments in time rather than over all data falling within a
map-pixel. This procedure results in ``pure'' $T$, $Q$ and $U$
time-streams for each detector but at a reduced data rate determined
by the number of data samples over which the averaging of
equation~(\ref{eqn:tqu_mapmaking}) is performed. For our second
analysis, we first difference each detector pair and then demodulate
the differenced time-streams using equation~(\ref{eqn:qu_mapmaking}),
again applied over short segments of time, resulting in ``pure'' $Q$
and $U$ time-streams for each detector pair, once again at a reduced
data rate. Maps of the $Q$ and $U$ Stokes parameters are then
constructed by simple binning of the demodulated $Q$ and $U$
time-streams from all detectors or detector pairs. In the case where
detector pairs are differenced, we are explicitly combining
information from multiple detectors to separate the temperature and
polarization signals whilst when we do not difference, we are
attempting to separate the $T$ and $Q/U$ signals present in detector
time-stream in isolation.

\begin{figure}
  \centering
  \resizebox{0.48\textwidth}{!}{  
    \rotatebox{-90}{\includegraphics{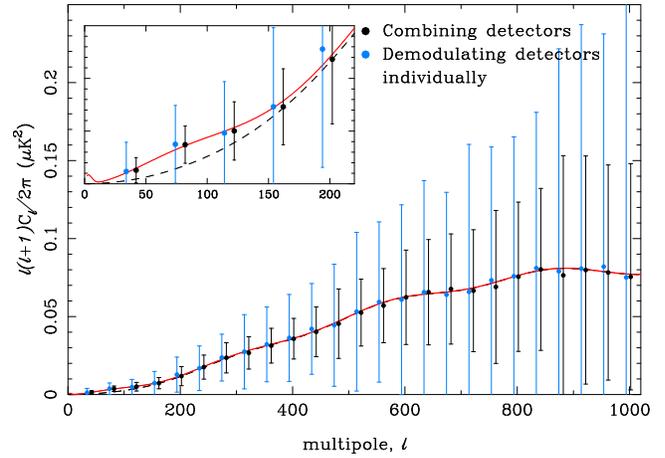}}}
  \caption{Comparison between the $B$-mode power spectra recovered
  using an analysis based on differencing detector pairs (black
  points) and one based on demodulating each detector individually
  (blue points). In the presence of a time-varying common-mode
  systematic, such as the $1/f$ atmospheric noise we have considered
  here, the analysis based on detector differencing is far superior to
  the analysis based on demodulating each detector individually.}
  \label{fig:cls_diff_nodiff}
\end{figure}

The results of these tests are shown in Figs.~\ref{fig:maps_diff_nodiff} and
\ref{fig:cls_diff_nodiff}. Figure~\ref{fig:maps_diff_nodiff} shows the noise-only
$U$-polarization maps recovered using the two different analysis
techniques. Although striping from the atmospheric fluctuations is not
present in either of the maps, the extra uncertainty introduced when
one attempts to separate the $T$ and $Q/U$ signals from individual
detectors in isolation clearly results in an increased white noise
level in the polarization maps. This results in a degradation factor
of $\sim 2$ in the resulting measurements of the $B$-mode power
spectrum on all angular scales (Fig.~\ref{fig:cls_diff_nodiff}) with
a corresponding degradation in the detection significances for both
a measurement of the total $B$-mode signal and for a detection of $r=0.026$
(Table~\ref{tab:diff_nodiff}). These results are in excellent
agreement with those of \cite{sutton09} who found it necessary to
apply optimal mapping techniques to rapidly modulated single-detector
time-streams in the presence of $1/f$ atmospheric noise. 

We emphasise that the results presented in
Fig.~\ref{fig:cls_diff_nodiff} and in Table~\ref{tab:diff_nodiff} for
the case where we have analysed single detectors in isolation would
likely be improved if the Fourier domain filtering used in
\cite{johnson07} was implemented. We are currently working to
integrate this step into our algorithm, and we expect to report the
subsequent improvement in future publications.
  
\begin{table}
\caption{Detection significances (in units of $\sigma$) from the
  analysis of identical simulated data with a HWP rotating
  continuously at $3$~Hz. The first analysis is based on detector
  differencing, the second based on demodulation of individual
  detectors in isolation.}
\begin{center}
\begin{tabular}{c|c|c|c}
Analysis & $E$-mode & $B$-mode & $r = 0.026$ \\
\hline
Detector differencing & $132.9$ & $10.14$  & $1.43$ \\
Demodulation          & $123.4$ & $5.11$   & $0.68$ \\
\hline
\end{tabular}
\end{center}
\label{tab:diff_nodiff}
\end{table}

\subsection{Comparison of simulated and predicted \clover\, performance}
\label{sec:fisher}
It is common practice to make predictions of the performance of
upcoming experiments using a Fisher-matrix analysis which
attempts to predict the achievable errors on, for example, power
spectra or cosmological parameters under some simplifying
assumptions. Generally these assumptions will include uniform coverage
of the observing fields and uncorrelated Gaussian noise resulting in
an isotropic and uniform white-noise distribution across the observing
field. In contrast, the work described in this paper has made use of a
detailed simulation pipeline which we have created for the \clover\,
experiment. Our simulation pipeline includes the \clover\,
focal-plane designs as well as a realistic scan strategy appropriate
for observing the four chosen \clover\, fields from the telescope site
in Chile. In addition we have employed a detailed model of the TES
detector noise properties and responsivity, and $1/f$ atmospheric
noise correlated across the focal-plane array.  Moreover, our errors
are calculated using a Monte-Carlo analysis and so should
automatically include any effects due to correlations between map
pixels etc. An interesting exercise therefore is to compare the
expected errors from a Fisher-matrix analysis to those obtained
from our simulation analysis. 

The polarization band-power Fisher matrix is (e.g.~\citealt{tegmark01})
\be
\mathrm{cov}^{-1}_{(bP)(bP)'} = \frac{1}{2}\mathrm{tr}\left(
\mC^{-1} \frac{\partial \mC}{\partial \mathcal{C}_b^P}
\mC^{-1} \frac{\partial \mC}{\partial \mathcal{C}_{b'}^{P'}}
\right)
\ee
where $P$ and $P'$ are $E$ or $B$, and $b$ labels the bandpower. Here,
$\mC$ is the covariance matrix of the noisy Stokes maps and
$\mathcal{C}_b^P$ are bandpowers of $\ell (\ell +1)C_\ell^P/(2\pi)$.
We analyse a single circular field with area equal to that retained
in the pseudo-$C_\ell$ analysis described in Section~\ref{subsec:power},
and multiply the Fisher matrix by four to account for the
number of fields observed (which are thus assumed to be fully independent).
We ignore inhomogeneity of the noise in the maps so that our problem
has azimuthal symmetry about the field centre. This allows us to
work in a basis where the data is Fourier transformed in azimuth and
the covariance matrix becomes block diagonal, thus speeding up the
computation of the Fisher matrix considerably. The Fisher matrix
takes full account of band-power correlations (both between $b$ and
polarization type) and the effect of ambiguities in isolating
$E$ and $B$-modes given the survey geometry. We deal with power
on scales larger than the survey by including a junk band-power for
each of $E$ and $B$ whose contribution to $\mathrm{cov}_{(bP)(bP)'}$
we remove before computing detection significances.

\begin{figure}
  \centering
  \resizebox{0.48\textwidth}{!}{  
    \rotatebox{-90}{\includegraphics{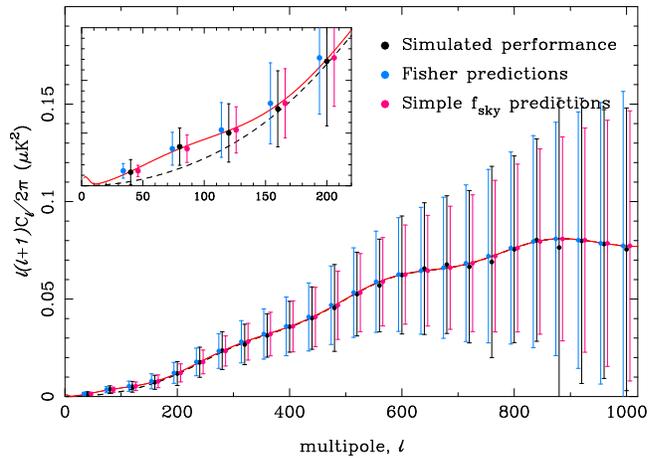}}}
  \caption{Comparison between the predicted performance of \clover\,
  as calculated using a Fisher-matrix analysis and the simulated
  performance from our Monte-Carlo pipeline (for our reference
  simulation). Also shown for comparison are the errors predicted from
  a na\"{\i}ve $f_{\rm sky}$ analysis.}
  \label{fig:clover_compare}
\end{figure}

The comparison between the predicted and simulated performance of
\clover\, is shown in Fig.~\ref{fig:clover_compare}. In this plot, we
also include the predicted $B$-mode errors from a na\"{\i}ve mode-counting
argument based on the fraction of sky observed, $f_{\rm sky}$. For
these estimates, we assume independent measurements of the power
spectrum in bands of width $\Delta \ell$ given by
\be
(\Delta C_\ell)^2 = \frac{2}{(2\ell + 1) f_{\rm sky} \Delta \ell}
(C_{\ell} + N_{\ell})^2,
\ee
where $C_\ell$ is the band-averaged input signal and $N_\ell$ is the
band-averaged noise. For uncorrelated and isotropic Gaussian random
noise, the latter is given by $N_\ell = w^{-1} B_\ell^{-2}$
where $B_\ell = \exp( - \ell (\ell + 1) \sigma_B^2 / 2)$ is the
transform of the beam with $\sigma_B = \theta_B / \sqrt{8 \ln 2}$ for
a beam with FWHM of $\theta_B$. The weight $w^{-1} =
\Omega_{\rm pix} \sigma^2_{\rm pix}$, where the pixel noise in the $Q$
and $U$ maps is 
\begin{equation}
\sigma^2_{\rm pix} = \frac{ ({\rm NET}/\sqrt{2})^2 \Theta^2}{t_{\rm obs}
  (N_{\rm det}/4) \Omega_{\rm pix}}. 
\label{eqn:pixel_noise}
\end{equation}
Here $\Theta^2$ is the total observed area, $t_{\rm obs}$ is
the total observation time and $\Omega_{\rm pix}$ is the pixel
size. In equation~(\ref{eqn:pixel_noise}), we have used ${\rm
  NET}/\sqrt{2}$ to account for the fact that a single measurement of
$Q$ or $U$ requires a measurement from two detectors (or,
alternatively, two measurements from a single detector) and we use
$N_{\rm det}/4$ as the effective number of $Q$ and $U$ detectors.

Over most of the $\ell$ range, the agreement between the Fisher matrix
predictions and the simulated performance is rather good -- the only
significant discrepancy is for the lowest band-power where the
simulations fail to match the predicted Fisher error. This is almost
certainly due to the relatively poor performance of our power spectrum
estimator on the very largest scales where pseudo-$C_\ell$ techniques
are known to be sub-optimal (compared to, for example, a maximum
likelihood analysis). In terms of a detection of the total $B$-mode
signal, the Fisher analysis predicts a detection for \clover\, of
$\sim 12.0\sigma$. For comparison, the na\"{\i}ve $f_{\rm sky}$
analysis predicts a $12.4\sigma$ detection. For our assumed
tensor-to-scalar ratio, the Fisher matrix analysis predicts $r=0.026
\pm 0.013$ (a $2.04\sigma$ detection) and the na\"{\i}ve analysis
yields $r=0.026 \pm 0.011$ (a $2.39\sigma$ detection). Comparing to
the detection significances quoted in Table~\ref{tab:simsummary}, we
see that the detections recovered from the simulations fail to match
these numbers. For the case of the total $B$-mode amplitude, this
discrepancy is entirely due to the fact that we are unable to measure
and include in our analysis (anti-)correlations between the
band-powers measured from our small number (50) of simulations --
when we neglect the correlations in the Fisher matrix analysis, the
Fisher prediction drops to a $10.24\sigma$ detection, in excellent
agreement with our measured value from simulations. For the primordial
$B$-mode signal only, the discrepancy found is also partly due to the
same effect (ignoring correlations in the Fisher analysis reduces the
Fisher prediction for primordial $B$-modes to $1.9\sigma$). As
mentioned above, we suspect that the additional decrease in sensitivity to
primordial $B$-modes seen in the simulations is due to the slightly
sub-optimal performance of our implementation of the pure
pseudo-$C_\ell$ estimator on the largest scales.

We should point out that in this work, we have made no attempt to optimise the
survey strategy in light of recent instrument developments. In
particular, the survey size we have adopted for these simulations was
optimised for a measurement of $r=0.01$ with \clover\, when the
experiment was expected to have twice the number of detectors now
planned. For the instrument parameters we have adopted in this analysis
(which are a fair representation of the currently envisaged
experiment), the optimal survey area for a measurement of $r=0.026$ 
would be significantly smaller than the $\sim 1500$ deg$^2$ we have
used here due to the increased noise levels from the reduced number of
detectors. Alternatively, if we had assumed a larger input value of
$r$, the optimal survey size would increase. Optimisation of both the
survey area and the scan strategy in light of these changes in the
instrument design is the subject of on-going work. 
There are, of course, many other sources of uncertainty which we have
not yet accounted for in our simulation pipeline and so both the
predicted and simulated performance numbers should be taken only as
guidelines at this time. However, it is encouraging that the extra
sources of uncertainty which are included in our simulation pipeline
(realistic instrument parameters, a realistic scan strategy,
correlated noise), in addition to any uncertainties introduced as part
of our subsequent analysis of the simulated data, do not degrade the
expected performance of \clover\, by a large amount. 

\section{Conclusions}
\label{sec:conclusions}
We have performed a detailed investigation of the ability of both slow
and fast polarization modulation schemes to mitigate possible
systematic effects in upcoming CMB polarization experiments, targeted
at measuring the $B$-mode signature of gravitational waves in the
early universe. To do this we have used a simulation pipeline
developed in the context of the \clover\, experiment, which includes
realistic instrument and observation parameters as well as $1/f$
detector noise and $1/f$ atmospheric noise correlated across the
\clover\, focal-plane array. Using this simulation tool, we have
performed simulations of \clover\, operating with no explicit
modulation, with a stepped HWP and with a HWP rotating continuously at
$3$~Hz. We have analysed the resulting time-stream simulations using
the technique of detector differencing coupled with a na\"{\i}ve map-making
scheme, and finally have reconstructed the $E$ and $B$-mode power
spectra using an implementation of the near-optimal ``pure''
pseudo-C$_\ell$ power spectrum estimator.

As expected, we find that fast modulation via a continuously rotating
HWP is extremely powerful in mitigating a correlated $1/f$ component
in the detector noise but that a stepped HWP is not. In addition, we
have demonstrated that a polarized $1/f$ component in the atmosphere
is not mitigated by any amount of modulation and if present, would
need to be mitigated in the analysis using a sophisticated map-making
technique. We have further verified with simulations that fast
modulation is very effective in mitigating instrumental polarization
that is fixed relative to the instrument basis, for example the $T
\rightarrow Q$ leakage caused by systematic gain errors and
mis-matches between detectors, in agreement with the conclusions
of~\citet{odea07}. We have also demonstrated that modulation does not
mitigate a systematic mis-calibration of polarization angles and that
these angles will need to be measured accurately in order to avoid a
systematic leakage between $E$ and $B$-modes. The other systematics
which we have investigated (pointing errors, mis-estimated
time-constants) have a negligible impact on the recovered power
spectra for the parameters adopted in our simulations.

In addition to our investigation of systematic effects, we have
stressed the importance of combining data from multiple detectors and
have demonstrated the superior performance of a differencing technique
as opposed to one based on measuring all three Stokes parameters from
single detectors in isolation. We suggest that the latter technique,
although possible in the presence of rapid modulation, is likely a
poor choice of analysis technique, at least in the presence of a
common-mode systematic effect such as atmospheric $1/f$ noise.

Finally, we have compared the simulated performance of the \clover\,
experiment with the expected performance from a simplified
Fisher-matrix analysis. For all but the very lowest multipoles, where
the simulations fail to match the Fisher predictions, we find
excellent agreement between the predicted and simulated
performance. In particular, despite the highly anisotropic noise
distribution present in our simulated maps, our measurement of the
total $B$-mode signal matches closely with the Fisher matrix
prediction (the latter assuming isotropic noise).  On the other hand,
the measurement of the large scale $B$-mode signal (and thus of the
tensor-to-scalar ratio, $r$) from the simulations is around 20 per
cent worse
than the Fisher prediction.  This is almost certainly due to the
sub-optimality of our power spectrum analysis on large scales. It is
possible that the Fisher matrix predictions could be recovered from
the simulations by using a more optimal weighting scheme in the pure
pseudo-$C_\ell$ analysis, or, more likely, by using a
maximum-likelihood $C_\ell$ estimator for the low multipoles.

One important class of systematic effects which we have not considered
in this paper are those associated with imperfect optics. Additionally, 
we have considered only two effects associated with an imperfect HWP.
The efficacy of fast modulation to mitigate systematic
effects from imperfect optics, for example instrumental polarization
due to beam mis-match, is expected to depend critically on the optical
design (such as the HWP location).  We are currently working to
include such optical systematics in our simulation pipeline, along
with a more detailed physical model of the atmosphere and models of
the expected polarized foreground emission. In future work, in
addition to investigating further systematic effects, we will extend
our simulations to multi-frequency observations and will use these to
test alternative foreground removal techniques. We will also apply the
``destriping'' map-making technique of \cite{sutton09} to our
simulations to assess the relative merits of destriping in analysis as
opposed to a hardware based approach for mitigating $1/f$ noise.
 
\section*{Acknowledgments}
We are grateful to the \clover\, collaboration for useful
discussions. We thank Kendrick Smith for making his original pure
pseudo-$C_\ell$ code available which we adapted to carry out some of
the analysis in this paper. We thank John Kovac and Jamie Hinderks for
the up-to-date descriptions of the {\sevensize BICEP-2/KECK} and \piper\,
experiments respectively. The simulation work described in this paper was carried
out on the University of Cambridge's distributed computing facility,
\camgrid. We acknowledge the use of the \fftw\, \citep{frigo05},
\camb\, \citep{lewis00} and \healpix\, \citep{gorski05} packages.

\setlength{\bibhang}{2.0em}

\appendix
\section{$E$-$B$ leakage from pointing jitter}
\label{app:pointing}

For the case of random pointing jitter and a scan/modulation
strategy that ensures every sky pixel is visited with a wide range
of polarization sensitivity directions, it is straightforward to estimate
the spurious $B$-mode power that is induced from $E$-modes for a single
focal-plane pixel. We work in the flat-sky approximation where
directions on the sky are denoted by positions $\vx$ in the tangent plane.
A given sky pixel $p$ receives $N_p$ hits with polarization sensitivity
directions $\{ \phi_i \}$, where $i=1,\ldots,N_p$, and assume these
angles are uniformly distributed.
If $d_i$ is the $i$th differential measurement from the detector pair
contributing to pixel $p$, then we can approximate the recovered
Stokes parameters in that pixel (equation~\ref{eqn:qu_mapmaking})
as
\begin{eqnarray}
\hat{Q}(\vx_p) &\approx& \frac{2}{N_p} \sum_i d_i \cos (2\phi_i) \nonumber \\
\hat{U}(\vx_p) &\approx& \frac{2}{N_p} \sum_i d_i \sin (2\phi_i) .
\end{eqnarray}
If the $i$th observation has a pointing error $\valpha_i$, then
the signal contribution to $\hat{Q}(\vx_p)$, for example, is
\begin{eqnarray}
\hat{Q}(\vx_p) &\approx& \frac{2}{N_p} \sum_i \left[Q(\vx_p+\valpha_i)
\cos^2(2\phi_i) \right. \nonumber \\
&&\mbox{} \left. \phantom{xxxxxxx}
+ U(\vx_p+\valpha_i)\sin(2\phi_i) \cos(2\phi_i)\right] .
\end{eqnarray}
Note that here, $Q$ and $U$ are the beam-smoothed fields.
Approximating the displaced Stokes parameters by a gradient approximation,
we find
\be
\hat{Q}(\vx_p) \approx Q(\vx_p) + \valpha_{\mathrm{eff}} \cdot \vgrad Q
|_{\vx_p}
+ \vbeta_{\mathrm{eff}} \cdot \vgrad U |_{\vx_p},
\label{eq:gradientQ}
\ee
where the effective angles
\begin{eqnarray}
\valpha_{\mathrm{eff}} &\equiv& \frac{2}{N_p} \sum_i \cos^2(2\phi_i)
\valpha_i \nonumber \\
\vbeta_{\mathrm{eff}} &\equiv& \frac{2}{N_p} \sum_i \cos(2\phi_i)\sin(2\phi_i)
\valpha_i .
\end{eqnarray}
Repeating for the recovered $U$ Stokes parameter, we find
\be
\hat{U}(\vx_p) \approx U(\vx_p) + \vgamma_{\mathrm{eff}} \cdot \vgrad U
|_{\vx_p} + \vbeta_{\mathrm{eff}} \cdot \vgrad Q |_{\vx_p}
\label{eq:gradientU}
\ee
where
\be
\vgamma_{\mathrm{eff}} \equiv \frac{2}{N_p} \sum_i \sin^2(2\phi_i)
\valpha_i .
\ee
Note that for pointing errors that are not constant in time,
$Q$ and $U$ are generally displaced differently and there is
non-local $Q$-$U$ coupling through $\vbeta_{\mathrm{eff}}$.
We recover the simple map-based model of pointing errors used
in~\citet{hu03} and~\citet{odea07},
\be
(\hat{Q}\pm i \hat{U})(\vx_p) \approx (Q\pm i U)(\vx_p) +
\valpha \cdot \vgrad (Q\pm i U)|_{\vx_p} ,
\ee
for the case of constant pointing errors, i.e.\ $\valpha_i = \valpha$.

In the limit that the pointing jitter is independent between time samples,
the pointing errors in the map, $\valpha_{\mathrm{eff}}$,
$\vbeta_{\mathrm{eff}}$ and $\vgamma_{\mathrm{eff}}$ are independent
between pixels. (Note that this will not hold if maps are made
from more than one focal-plane pixel.) These pointing errors are
mean zero, and have independent $x$- and $y$-components. The
non-vanishing correlators are
\begin{eqnarray}
\langle \valpha_{\mathrm{eff}} \cdot \valpha_{\mathrm{eff}}
\rangle &\approx& 3 \alpha_{\mathrm{tod}}^2 / (2N_p) \\
\langle \vbeta_{\mathrm{eff}} \cdot \vbeta_{\mathrm{eff}}
\rangle &\approx& \alpha_{\mathrm{tod}}^2 / (2N_p) \\
\langle \vgamma_{\mathrm{eff}} \cdot \vgamma_{\mathrm{eff}}
\rangle &\approx& 3 \alpha_{\mathrm{tod}}^2 / (2N_p) \\
\langle \valpha_{\mathrm{eff}} \cdot \vgamma_{\mathrm{eff}}
\rangle &\approx& \alpha_{\mathrm{tod}}^2 / (2N_p) .
\end{eqnarray}
Here, $\alpha_{\mathrm{tod}}^2 = \langle \valpha_i \cdot \valpha_i \rangle$
is the variance of the pointing jitter, and we have approximated
discrete sums over the angles $\phi_i$ by their integrals.
We shall further assume that the
number of hits per pixel is uniform over the map in which case
the map-based pointing errors have homogeneous statistics through the map.
The quantity $\alpha_{\mathrm{tod}}^2 \Omega_{\mathrm{pix}} / N_p$
will arise in our final result for the spurious $B$-mode power,
where $\Omega_{\mathrm{pix}}$ is the pixel area. This product is independent
of pixel area since $N_p \propto \Omega_{\mathrm{pix}}$.
For 30-arcsec jitter that is random between 100~Hz samples,
$\alpha_{\mathrm{tod}} \sqrt{\Omega_{\mathrm{pix}}/N_p}
\approx 3\times 10^{-8}\,\mathrm{rad}^2$ for an
eight-hour observation of a 380-$\mathrm{deg}^2$ field.

The calculation of the $B$-mode power due to leakage of
$E$-modes follows from equations~(\ref{eq:gradientQ}) and
(\ref{eq:gradientU}) with the standard techniques used, for example,
in the calculation of gravitational lensing on the CMB
(see, e.g.~\citealt{lewis06}). We shall not give the details here
but simply note that the final result is a white-noise contribution
with
\be
C_\ell^B \approx \frac{\alpha_{\mathrm{tod}}^2\Omega_{\mathrm{pix}}}{2N_p}
\int \frac{\ell^3 d\ell}{2\pi} C_\ell^E
e^{-\ell^2 \sigma_B^2}, 
\ee
where $\sigma_B$ is the beam size.
The integral here approximates the mean-squared gradient of the beam-smoothed
polarization which, for a $5.5$-arcmin beam
$\approx 1.8\times 10^7\,\mu\mathrm{K}^2$.
For our simulation parameters we therefore expect
$C_\ell^B <  10^{-8}\, \mu\mathrm{K}^2$
from pointing jitter which is $< 1$ per cent of
the large-scale power induced by gravitational lensing.

\label{lastpage}

\end{document}